\documentclass[aps,pra,twocolumn,amsmath,amssymb,nofootinbib,showpacs,superscriptaddress,longbibliography]{revtex4-1}
\usepackage[english]{babel}
\usepackage{latexsym}
\usepackage{graphics}
\usepackage{graphicx}
\usepackage{epsfig}
\usepackage{color}
\usepackage{bm}
\usepackage{amsmath}
\usepackage{amssymb}
\usepackage{amsthm}
\usepackage{dcolumn}
\usepackage{bm}
\usepackage{float}
\usepackage{hyperref}
\usepackage{color}
\usepackage{epstopdf}
\usepackage{braket}
\usepackage{cleveref}
\usepackage{braket, comment}
\usepackage[svgnames]{xcolor}
\usepackage{listings}
\usepackage{bbold}
\hypersetup{hidelinks,colorlinks=true,allcolors=DarkBlue}

\newcommand{\BLUE}[1]{{\color{black}#1}}

\theoremstyle{remark}

\definecolor{codegreen}{rgb}{0,0.6,0}
\definecolor{codegray}{rgb}{0.5,0.5,0.5}
\definecolor{codepurple}{rgb}{0.58,0,0.82}
\definecolor{backcolour}{rgb}{0.98,0.98,0.98}

\lstdefinestyle{mystyle}{
    backgroundcolor=\color{backcolour}, 
    commentstyle=\color{codegreen},
    keywordstyle=\color{magenta},
    numberstyle=\tiny\color{codegray},
    stringstyle=\color{codepurple},
    basicstyle=\ttfamily\footnotesize,
    breakatwhitespace=false,         
    breaklines=true,                 
    captionpos=b,                    
    showspaces=false,                
    showstringspaces=false,
    showtabs=false,                  
    tabsize=2
}

\begin{document}

\title{Exploring postselection-induced quantum phenomena\\with time-bidirectional state formalism}

\author{Evgeniy O. Kiktenko}
\affiliation{Department of Mathematical Methods for Quantum Technologies, Steklov Mathematical Institute of Russian Academy of Sciences, Moscow 119991, Russia}
\affiliation{Russian Quantum Center, Skolkovo, Moscow 121205, Russia}
\affiliation{Geoelectromagnetic Research Center, Schmidt Institute of Physics of the Earth, Russian Academy of Sciences, Troitsk 108840, Russia}
\affiliation{National University of Science and Technology “MISIS”, Moscow 119049, Russia}
\email{e.kiktenko@rqc.ru}

\date{\today}
\begin{abstract}
Here we present the time-bidirectional state formalism (TBSF) unifying in a general manner the standard quantum mechanical formalism with no postselection and the time-symmetrized two-state (density) vector formalism, which deals with postselected states.
In the proposed approach, a quantum particle's state, called a time-bidirectional state, is equivalent to a joined state of two particles propagating in opposite time directions.
For a general time-bidirectional state, we derive outcome probabilities of generalized measurements, as well as mean and weak values of Hermitian observables.
We also show how the obtained expressions reduce to known ones in the special cases of no postselection and generalized two-state (density) vectors.
Then we develop tomography protocols based on mutually unbiased bases and a symmetric informationally complete positive operator-valued measure, allowing experimental reconstruction of an unknown single qubit time-bidirectional state.
Finally, we employ the developed techniques for tracking of a qubit's time-reversal journey in a quantum teleportation protocol realized with a cloud-accessible noisy superconducting quantum processor.
The obtained results justify an existence of a postselection-induced qubit's proper time-arrow, which is different from the time-arrow of a classical observer, and demonstrate capabilities of the TBSF for exploring quantum phenomena brought forth by a postselection in the presence of noise.
\end{abstract}

\maketitle

\section{Introduction}

\begin{figure}
    \centering
    \includegraphics[width=\linewidth]{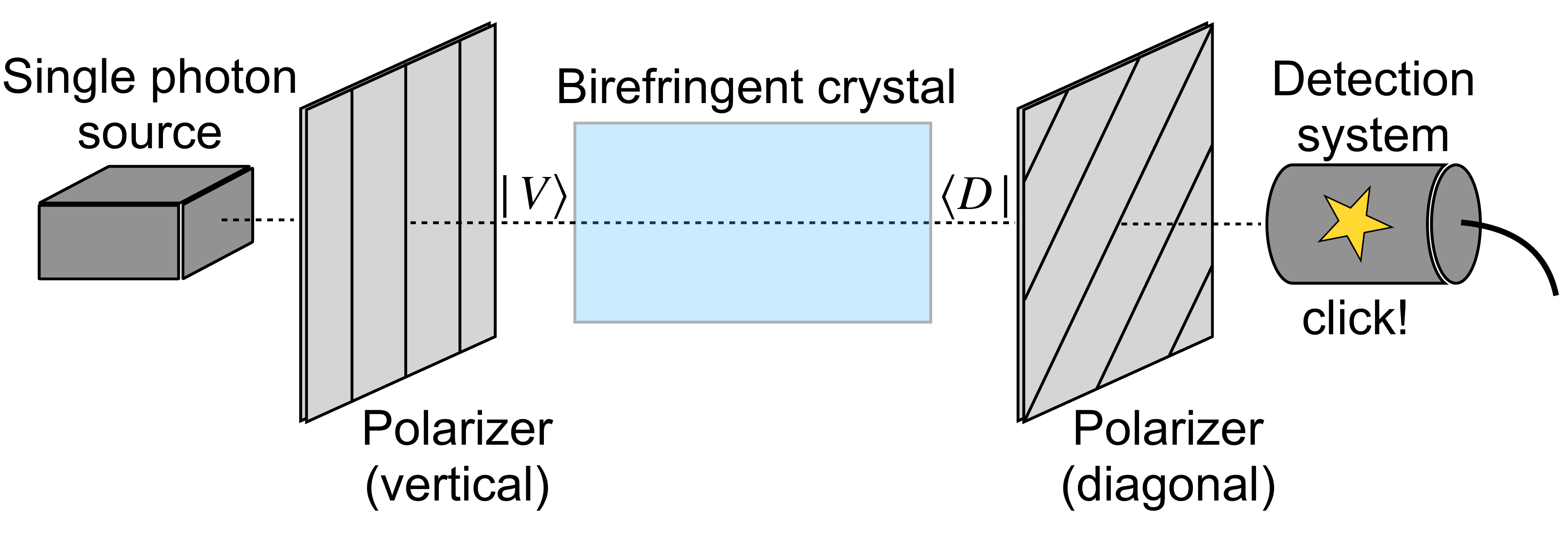}
    \caption{
    \BLUE{
    An example of an optical experimental setup dealing with both pre- and postselected states.
    Here photons pass through two polarizers, the first of which filters out vertically polarized states $\ket{V}$ and the second one filters out diagonally polarized states $\ket{D}$ ($|\bra{V}H\rangle|=2^{-1/2}$).
    By introducing a two-state vector $(\ket{V}, \bra{D})$, the TSVF provides a complete description of photons' polarization, given their detection after the second polarizer (the postselection condition).
    In particular, the two-state vector allows describing a transformation of photons' spatial degree of freedom due to a coupling with polarization realized by a birefringent crystal~(see a comprehensive discussion in Ref.~\cite{dressel2014colloquium}).}
    }
    \label{fig:optical-scheme}
\end{figure}

The standard quantum formalism is commonly used for calculating a probability distribution of measurement outcomes, given a complete characterization of preparation and evolution of a measured quantum system.
%about the measured quantum system's preparation and an evolution map acting between the preparation and the measurement.
This consideration with respect to a preselected initial state contains an implicit time asymmetry related to the concept of `collapsing', or `reduction', of system's state in the measurement process~\cite{watanabe1955symmetry, aharonov1964time}. %The initial state is assumed to be known, i.e. preselected, while the measurement outcome is generally uncertain
Combining the preselection with a postselection, i.e. the consideration of a particular outcome of the measurement, removes this asymmetry and gives rise to the two-state vector formalism (TSVF)~\cite{aharonov1964time, aharonov2008two}.
\BLUE{Within the TSVF, the quantum particle's state is described by a pair of vectors $(\ket{\psi},\bra{\phi})$, where $\ket{\psi}$, determined by the preselection, can be considered evolving forward in time, while $\bra{\phi}$, determined by the postselection, can be considered as evolving back from the future to the past
(see an example of an optical experiment with pre- and postselection in Fig.~\ref{fig:optical-scheme})}. 

From a practical point of view, one of the most important concepts appearing in the framework of postselection and the TSVF is weak values of observables~\cite{aharonov1988result,aharonov1990properties}.
Despite some criticism (see e.g.~\cite{leggett1989comment, peres1989quantum, aharonov1989aharonov}), weak values and related techniques for a weak value amplification~\cite{pang2014entanglement, jordan2014technical, harris2017weak, xu2020approaching} appear to be extremely useful in the context of quantum metrology~\cite{hosten2008observation,dixon2009ultrasensitive, strubi2013measuring, xu2013phase, zhou2013weak, jayaswal2014observing, magana2014amplification, lyons2015power, hallaji2017weak,arvidsson2020quantum} (for a review, see~\cite{dressel2014colloquium}).
Moreover, taking postselection into account also plays an important role in studying complexity theory~\cite{aaronson2005quantum}, quantum contextuality~\cite{leifer2005pre, tollaksen2007pre, pusey2014anomalous, pusey2015logical, kunjwal2019anomalous}, fundamentals of quantum physics~\cite{aharonov2002revisiting, dressel2010contextual, kocsis2011observing, lundeen2011direct, danan2013asking, vaidman2013past, vaidman2014tracing, vaidman2017weak, xu2019measurements, cimini2020anomalous, rebufello2021anomalous}, design of quantum computing algorithms~\cite{harrow2009quantum}, and quantum communication protocols~\cite{arvidsson2016quantum, arvidsson2017evaluation, arvidsson2019postselection, vaidman2019analysis, wander2021three}.

The postselection with respect to entangled states gives rise to time-reversal phenomena, including an appearance of closed timelike curves (CTC), considered both theoretically~\cite{vaidman2007backward, laforest2006time, aharonov2009multiple, coecke2010quantum, svetlichny2011time, lloyd2011quantum, lloyd2011closed, korotaev2015quantum, oreshkov2015operational, shepelin2021multiworld} and experimentally~\cite{laforest2006time, lloyd2011closed}.
The basic idea behind these phenomena is that Bell state preparation and Bell state measurement can be considered as a kind of ``time mirrors'' reflecting a quantum state's propagation in time.
Note that this interpretation perfectly agrees with experimental results on delayed entanglement swapping~\cite{jennewein2001experimental, megidish2013entanglement}. 

Originally, the TSVF was formulated with respect to a pair of pure states~\cite{aharonov1964time, aharonov2008two}.
An important extension comes with introducing an ancillary particle and performing postselection with respect to an entangled state. 
This creates an entanglement between forward and backward evolving states of a two-state vector, and yields the concept of a generalized two-state vector~\cite{aharonov1991complete,aharonov2008two}.
Studying statistical ensembles of generalized two-state vectors bring forth a notion of two-state density vectors~\cite{silva2014pre}, which can be considered as a manifestation of density matrices in the framework of the TSVF.
Another approach of introducing mixedness into the TSVF is presented in Ref.~\cite{vaidman2017weak}, where the case of forward and backward evolving states described with density matrices is considered.

\begin{figure}
    \centering
    \includegraphics[width=\linewidth]{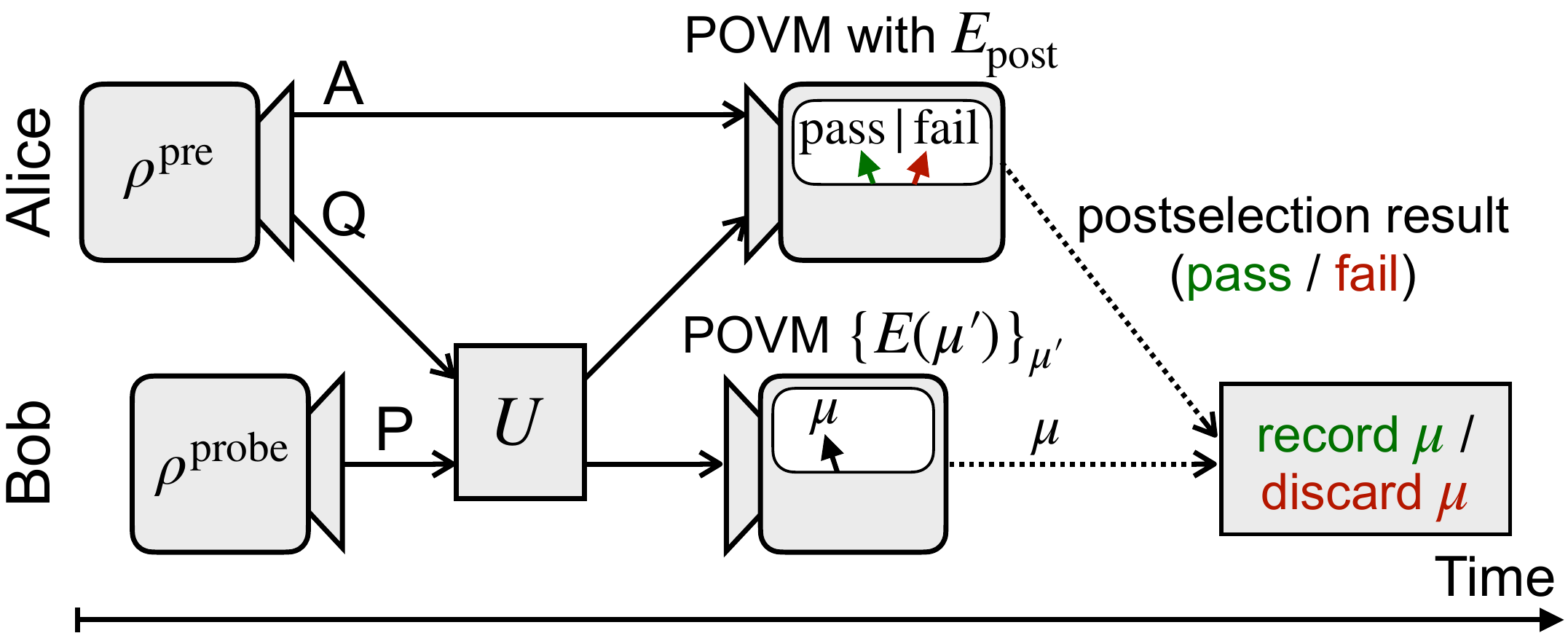}
    \caption{General scheme of a postselection experiment giving rise to a concept of a time-bidirectional state.
    First, Alice prepares particles Q and A in some arbitrary state and sends Q to Bob.
    On his side, Bob applies a unitary operation $U$ to Q and P, then performs an arbitrary measurement on P, and finally returns Q back to Alice.
    Then Alice makes a joint measurement on A and Q, described by a POVM containing a particular effect $E_{\rm post}$.
    If the outcome given by $E_{\rm post}$ is realized, then Alice tells Bob to keep his measurement result $\mu$, otherwise $\mu$ is discarded.
    }
    \label{fig:ps_scheme}
\end{figure}

The present work is devoted to a further development of effective description of quantum states in the presence of postselection and pursues the following two main goals.
The first goal is closing the gap between the previous approaches for describing mixed, or randomized, quantum states in the presence of the postselection and the standard quantum formalism without any postselection at all.
This goal is achieved by extending the two-state density vector formalism~\cite{silva2014pre} with a more general time-bidirectional state formalism (TBSF), where the postselection is performed with respect to an arbitrary positive operator-valued measure (POVM) effect (see Fig.~\ref{fig:ps_scheme}).
Within the the developed formalism, a state of a particle is described by a bipartite, generally mixed, state, called a time-bidirectional state, which is equivalent to a joint state of two particles propagating in opposite time-directions. 
We show that in the absence of a postselection, i.e. identity postselection effect, the backward evolving part appears to be in the maximally mixed state, while the forward evolving one coincides with a density matrix from the standard formalism.
An important feature of the TBSF is its ability to account for any kind of decoherence noise, affecting both pre- and postselection.

The second goal is developing practical schemes for tomography of both pre- and postselected states.
In the current work, we focus on the case of a single qubit that can be easily generalized to a multiqubit one.
Compared to a high level recipe for making a complete set of Kraus operators, sufficient for reconstructing an unknown pre- and postselected state, presented in Ref.~\cite{silva2014pre}, here we obtain explicit circuits for running tomography protocols on an arbitrary quantum processor.
For this purpose, we borrow two basic approaches for single-qubit tomography: the one based on mutually unbiased bases (MUBs) corresponding to measuring three components of a Bloch vector, and the second based on a symmetric informationally complete POVM (SIC-POVM) allowing reconstruction of unknown state with a measurement of a single type.

To demonstrate capabilities of the TBSF and developed tomography techniques, we consider a well-known time-reversal phenomenon appearing in a quantum teleportation protocol~\cite{laforest2006time, coecke2010quantum, lloyd2011closed}.
Namely, we track propagation of a qubit state, initially prepared by Alice, (i) forward in time to the moment of a Bell measurement on her qubit and a qubit from a pre-shared Bell pair, then (ii) back in time on Alice' qubit from the Bell pair to the moment of the Bell pair birth, and then (iii) forward in time on Bob's particle from the Bell pair.
For this purpose we use a seven-qubit cloud-accessible noisy superconducting quantum processor provided by IBM.
Although, experiments on the observation of a postselection-induced time-travel in quantum teleportation were considered previously~\cite{laforest2006time,lloyd2011closed}, to the best of our knowledge, this is the first time where it is demonstrated, using the developed formalism, how the state, prepared by Alice, propagates back in time on Alice's physical qubit taken from the pre-shared Bell pair.
As already mentioned, an important advantage of the developed TBSF, compared, e.g., to the time-reversal formalism suggested in Ref.~\cite{laforest2006time}, is that this formalism allows considering decoherence effects.
In particular, we observe evidences of irreversible corruption of the Alice's quantum state during propagation along its proper postselection-induced time-arrow.

The rest of the paper is organized as follows.
In Sec.~\ref{sec:twotimetensors}, we introduce the concept of time-bidirectional states, provide some illustrative examples, and derive their main mathematical properties.
In Sec.~\ref{sec:measurements}, we discuss description of measurements made on a time-bidirectional state with a focus on von Neuman measurement of Hermitian observables and measurements of weak values.
In Sec.~\ref{sec:tomography}, we develop tomography protocols for experimental reconstructing of an unknown single-qubit time-bidirectional state.
In Sec.~\ref{sec:experiment}, we apply the developed formalism and tomography techniques for observing a time-reversal journey of a qubit's state in a quantum teleportation protocol.
We conclude and provide an outlook in Sec.~\ref{sec:concl}.

\section{Introducing time-bidirectional states} \label{sec:twotimetensors}

\subsection{General postselection experiment} \label{sec:postsel-scheme}

Let us consider a postselection experiment shown in Fig.~\ref{fig:ps_scheme}.
The experiment is realized by two parties, named Alice and Bob, that are able to communicate with quantum particles and classical messages.
%The first party, named Alice, is responsible for state preparation and postselection, while the second party, Bob, performs generalized indirect measurement on the state pre- and postselected by Alice.
At the start of the experiment, Alice prepares two particles, Q and A, in an arbitrary joint mixed state
\begin{equation}
    \rho^{\rm pre} = \rho^{\rm pre}_{ii';mm'}\ket{i}_{\rm Q}\bra{i'} \otimes \ket{m}_{\rm A}\bra{m'}.
\end{equation}
Here $\ket{n}_X$ with integer labels $n$ denote computational basis states of particle X and $\rho^{\rm pre}_{ii';mm'}$ are density matrix elements providing standard conditions $\rho^{\rm pre}\geq 0$, ${\rm Tr}\rho^{\rm pre}=1$.
Note that here and hereafter we apply the Einstein summation convention and omit explicit summation with respect to repeated indices.
After its preparation, Q is given to Bob, while A remains with Alice.

On his side, Bob takes an additional particle P, prepared in some mixed state
\begin{equation}
    \rho^{\rm probe} = \rho^{\rm probe}_{kk'} \ket{k}_{\rm P}\bra{k'}
\end{equation}
($\rho^{\rm probe}\geq 0$, ${\rm Tr}\rho^{\rm probe}=1$), and lets P and Q evolve according to a unitary evolution operator
\begin{equation}
    U = U_{l;j}^{k;i} \ket{l}_{\rm P}\bra{k} \otimes \ket{j}_{\rm Q}\bra{i}.
\end{equation}
Then Bob makes a measurement on P described by a POVM $\{E(\mu')\}_{\mu'}$, satisfying standard conditions $E(\mu') \geq 0$, $\sum_{\mu'} E(\mu')=\mathbb{1}$, where $\mathbb{1}$ is the identity matrix.
Here we consider outcome labels $\mu'$ belonging to an arbitrary finite set.
Bob keeps the obtained measurement outcome $\mu$, corresponding to the realized effect $E(\mu)$, and returns Q back to Alice.

On her side, Alice makes a joint measurement on Q and A, described by another POVM, whose collection of effects includes a particular effect
\begin{equation}
    E_{\rm post} = E_{\rm post}^{jj';mm'} \ket{j'}_{\rm Q}\bra{j} \otimes \ket{m'}_{\rm A}\bra{m}
\end{equation}
($0\leq E_{\rm post} \leq \mathbb{1}$).
\BLUE{If an outcome of Alice's measurement corresponds to $E_{\rm post}$, then we say that the postselection passed and set a special flag variable ${\sf ps}:=1$; postselection failed and ${\sf ps}:=0$ otherwise.}
Alice transmits the result of the postselection, i.e. single bit ${\sf ps}$, to Bob, who keeps his measurement outcome $\mu$ if postselection has passed, or discards $\mu$ otherwise.
We note that the only constraints on particular time moments, when the described operations take place, correspond to the general ordering:
Preparations of $\rho^{\rm pre}$ and $\rho^{\rm probe}$ are in the past light cone of $U$,  while Alice's and Bob's measurements are in the future light cone of $U$ and in the past light cone of the final decision on $\mu$.

\begin{figure*}
    \centering
    \includegraphics[width=\linewidth]{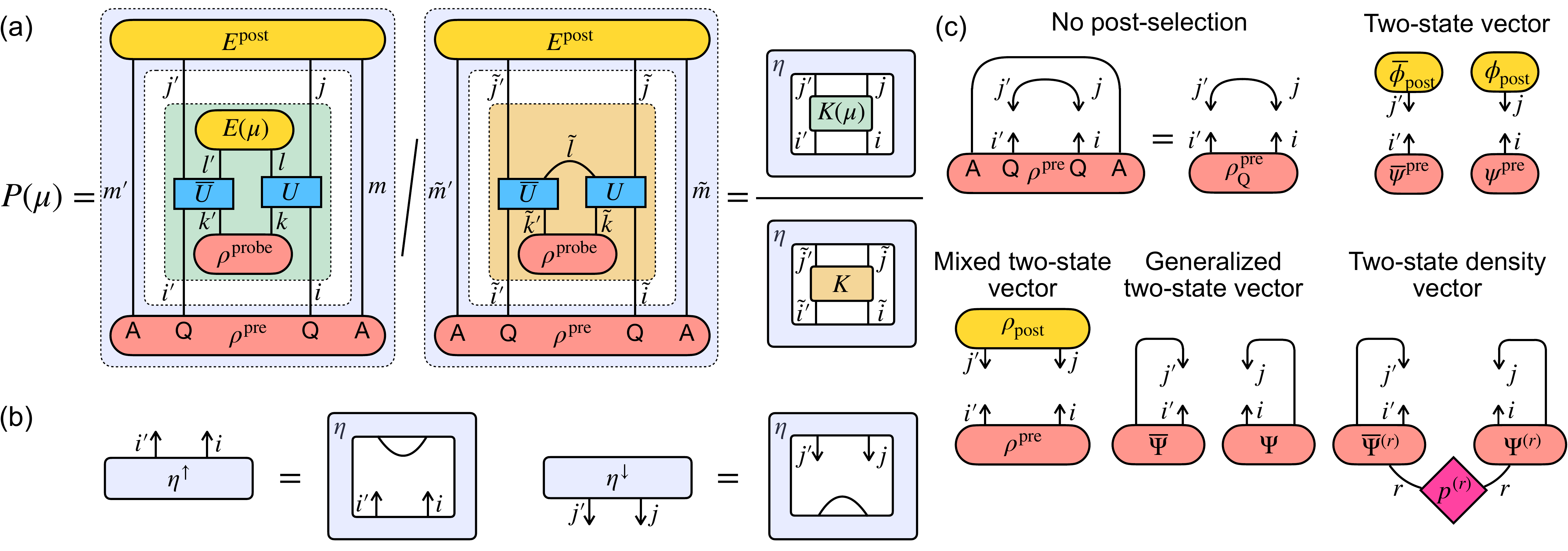}
    \caption{In (a) tensor network representation of Eq.~\eqref{eq:postsel-prob-huge} for $P(\mu)$, as well as definitions of time-bidirectional state $\eta$, operation outcome tensor $K(\mu)$, and operation tensor $K$ are shown.
    In (b) construction of reduced forward- and backward evolving reduced tensors $\eta^{\uparrow}$ and  $\eta_{\downarrow}$ is depicted.
    In (c) some examples of previously considered special types of time-bidirectional states are presented.
    }
    \label{fig:tensor-states}
\end{figure*}

\BLUE{
The main object of our study is a conditional probability distribution of Bob's measurement outcomes, given the passing postselection on Alice's side (here ${\cal M}$ denotes a random variable of Bob's outcome).
According to axiomatics of quantum mechanics, the probability of the joint event of ${\cal M}=\mu$ and the postselection passing is given by
\begin{multline}
    \Pr[{\cal M}=\mu \cup {\sf ps}=1]\\=\rho^{\rm pre}_{ii';kk'}\rho^{\rm probe}_{mm'}U_{l;j}^{i;m}\overline{U}_{l';j'}^{i';m'}E(\mu)^{ll'}E_{\rm post}^{jj';kk'},
\end{multline}
where the overbar denotes the complex conjugate.
The probability of the postselection passing takes the form
\begin{equation} \label{eq:postsel-prob-huge}
    \Pr[{\sf ps}=1] = \rho^{\rm pre}_{\tilde i \tilde i';\tilde k \tilde k'}\rho^{\rm probe}_{\tilde m m'}U_{\tilde l; \tilde j}^{\tilde i; \tilde m}\overline{U}_{\tilde l; \tilde j'}^{\tilde i';\tilde m'}E_{\rm post}^{\tilde j\tilde j';\tilde k\tilde k'} \equiv P_{\rm post}.
\end{equation}
Putting these expressions in Bayes' rule leads to
\begin{multline} \label{eq:prob-huge}
    %\Pr[{\cal M}=\mu | {\sf ps}=1]
    \Pr[{\cal M}=\mu|{\sf ps}=1]=
    \frac{\Pr[{\cal M}=\mu \cup {\sf ps}=1]}{P_{\rm post}}
    \\= 
    \frac{\rho^{\rm pre}_{ii';kk'}\rho^{\rm probe}_{mm'}U_{l;j}^{i;m}\overline{U}_{l';j'}^{i';m'}E(\mu)^{ll'}E_{\rm post}^{jj';kk'}}{
    \rho^{\rm pre}_{\tilde i \tilde i';\tilde k \tilde k'}\rho^{\rm probe}_{\tilde m m'}U_{\tilde l; \tilde j}^{\tilde i; \tilde m}\overline{U}_{\tilde l; \tilde j'}^{\tilde i';\tilde m'}E_{\rm post}^{\tilde j\tilde j';\tilde k\tilde k'}} \equiv P(\mu),
\end{multline}
given $P_{\rm post}>0$, and $P(\mu)=0$ otherwise.
It appears that Eq.~\eqref{eq:prob-huge} is much easier to follow in the form of a tensor network, shown in Fig.~\ref{fig:tensor-states}(a).
We note that if $Z$ is a tensor, then by $\overline{Z}$ we denote a tensor obtained from $Z$ by taking complex conjugation of all element.
It differs from the Hermitian conjugate of $Z$, which we denote as $Z^\dagger$.}

Within Eq.~\eqref{eq:prob-huge} we can separate two mathematical structures, which are related to Alice's and Bob's actions correspondingly.
The pre- and postselection, which are performed by Alice, are described by a tensor
\begin{equation}
    \eta_{ii'}^{jj'} := \rho^{\rm pre}_{ii';mm'} E_{\rm post}^{jj';mm'},
\end{equation}
while the indirect Bob's measurement is described by a collection of tensors
\begin{equation}
    K(\mu)^{ii'}_{jj'} := \rho^{\rm probe}_{kk'} U^{k;i}_{l;j}\overline{U}^{k';i'}_{l';j'}E(\mu)^{ll'}.
\end{equation}
We then can rewrite Eqs.~\eqref{eq:prob-huge} and~\eqref{eq:postsel-prob-huge} in a compact form
\begin{equation}
    \begin{aligned}
        P(\mu)&=\frac{K(\mu)_{mm'}^{ii'}\eta_{ii'}^{mm'}}{K_{\tilde m \tilde  m'}^{\tilde  i\tilde i}\eta_{\tilde i\tilde i'}^{\tilde m\tilde m'}} \equiv 
    \frac{K(\mu)\bullet\eta}{K\bullet\eta},\\
    \quad P_{\rm post}&=K\bullet\eta,
    \end{aligned}
\end{equation}
where $K:=\sum_\mu K(\mu)$ and $\bullet$, in line with Ref.~\cite{silva2014pre}, denotes contraction with respect to proper indices [see also Fig.~\ref{fig:tensor-states}(a)].

In what follows, we refer to $\eta$, $K(\mu)$, and $K$ as a time-bidirectional state, operation outcome $\mu$ tensor, and operation tensor correspondingly.
\BLUE{The time-bidirectional state $\eta$ describes the pre- and postselected state of the particle Q realized by assistance of initial entanglement and joint measurement of Q with the ancillary particle A. 
Operation outcome $\mu$ tensor $K(\mu)$ and operation tensors $K$ describe indirect measurement of Q realized by the probe particle P. 
}

One can see that $P(\mu)$ is invariant under multiplying $\eta$ (as well as both $K(\mu)$ and $K$) by a constant. 
So we have a freedom of renormalizing $\eta$ without affecting any of the observable quantities.
In our work we apply normalization in the form
\begin{equation} \label{eq:normalization}
    \eta \mapsto \frac{\eta}{\eta_{ii}^{jj}}.
\end{equation}
This normalization condition makes, as we see later, $\eta$ equivalent to a standard `preselected' joint state of two particles.
We note that $\eta_{ii}^{jj}$ is proportional to the probability of postselection given that Bob returns Q in the maximally mixed state.
So $\eta_{ii}^{jj}=0$ only in the trivial case, where the postselection probability $P_{\rm post}$ is 0 regardless of Bob's actions.

We can also introduce two `reduced' tensors
\begin{equation}
    \eta^{\uparrow}_{ii'}=\eta_{ii'}^{kk}, \quad \eta_{\downarrow}^{jj'}=\eta_{kk}^{jj'},
\end{equation}
which we call a `forward evolving' and `backward evolving' reduced tensors correspondingly [see Fig.~\ref{fig:tensor-states}(b)].
One can see that the semi-positivity of $\rho^{\rm pre}$ and $E_{\rm post}$ implies semi-positivity of $\eta^{\uparrow}$ and $\eta_\downarrow$. 
Provided normalization condition~\eqref{eq:normalization}, we also have ${\rm Tr}\eta^{\uparrow}={\rm Tr}\eta_{\downarrow}=1$.

\BLUE{As we will see, from the physical point of view, $\eta$ describes a joint state of two copies of the same degree of freedom considered for Q.
For example, if Q is a two-dimensional polarization degree of freedom of a photon, then $\eta$ is a $4\times4$ density matrix describing polarization of two photons taken from two distinct modes.
A seemingly confusing doubling of the state space appears due to taking into account both pre- and postselection.
Roughly speaking, the state of the first photon, given by $\eta^\uparrow$, is the state in which Q is prepared, and the state of the second photon, given by $\eta_\downarrow$, is the state in which Q is postselected (like $\ket{V}$ and $\bra{D}$ in Fig.~\ref{fig:optical-scheme}). 
Doing a postselection measurement on Q together with the ancillary particle A, initially entangled with Q, one can make the joint state $\eta$ of these photons entangled.
The question of how the joint state $\eta$ can be reconstructed is considered in detail in Sec.~\ref{sec:tomography}.}

%%% CORRECTIONS FROM THE REDLINE UP TO HERE

\subsection{Special cases of time-bidirectional states}

Here we consider some illustrative special cases of $\eta$ to get intuition behind $\eta^{\uparrow}$ and $\eta^{\downarrow}$.
Tensor diagrams for all the considered cases are shown in Fig.~\ref{fig:tensor-states}(c).

First, let us remove the postselection condition by taking $E_{\rm post}=\mathbb{1}$, which corresponds to a guaranteed passing of the postselection ($P_{\rm post}=1$).
In this case we have
\begin{equation}
    \eta_{ii'}^{jj'} = \rho^{\rm pre}_{ii';kk'} \delta_{jj'}\delta^{kk'},
\end{equation}
and after applying normalization condition~\eqref{eq:normalization}, $\eta$ factorizes into
\begin{equation} \label{eq:preselectedonly_state}
    \eta=\eta^{\uparrow}\otimes\eta_{\downarrow}, \quad \eta^{\uparrow}=\rho^{\rm pre}_{\rm Q}, \quad \eta_\downarrow= \rho_{\rm mix},
\end{equation}
where $\delta$ denotes a standard Kronecker symbol, $\rho^{\rm pre}_{\rm Q} = {\rm Tr}_{\rm A}\rho^{\rm pre}$ is a reduced preselected state of Q (here ${\rm Tr}_{\rm A}$ denotes a partial trace with respect to A), and $\rho_{\rm mix}\propto \mathbb{1}$ is the maximally mixed state.
The time-bidirectional state of the form~\eqref{eq:preselectedonly_state} has a clear physical meaning of the preselected state $\rho^{\rm pre}_{\rm Q}$ evolving forward in time, and a total uncertainty of the particle's future.
The probability to obtain $\mu$ on Bob's side takes a familiar form
\begin{equation}
\begin{aligned}
    P(\mu)&= \frac{\rho^{\rm pre}_{{\rm Q},ii'}\rho^{\rm probe}_{mm'}U_{l;j}^{i;j}\overline{U}_{l';j}^{i';m'}E(\mu)^{ll'}}{
    \rho^{\rm pre}_{{\rm Q}\tilde i \tilde i'}\rho^{\rm probe}_{\tilde m m'}U_{\tilde l; \tilde j}^{\tilde i; \tilde m}\overline{U}_{\tilde l; \tilde j}^{\tilde i';\tilde m'}}\\
    &={\rm Tr}\left[E(\mu){\rm Tr}_{\rm Q}\left(U\rho^{\rm pre}_{\rm Q}\otimes\rho^{\rm probe} U^\dagger\right)\right].
\end{aligned}
\end{equation}
This is exactly the value that one obtains with the use of the standard formalism.

The second illustrative special case is when $\rho^{\rm pre}$ is a product state of the pure state $\ket{\psi^{\rm pre}}=\sum_{i} \psi^{\rm pre}_i\ket{i}$ on Q and an arbitrary state of A, while postselection is realized only for Q with respect to the state $\bra{\phi_{\rm post}}=\sum_{i} \phi_{\rm post}^i\bra{i}$ ($E_{\rm post}\propto\ket{\phi_{\rm post}}_{\rm Q}\bra{\phi_{\rm post}} \otimes \mathbb{1}_{\rm A}$).
%is projector in the space of Q with respect to some state $\bra{\phi_{\rm post}}=\sum_{i} \phi_{\rm post}^i\bra{i}$ and identity in the space of A.
Elements of $\eta$ then take the form
\begin{equation}
    \eta_{ii}^{jj'}=\psi^{\rm pre}_{i}\overline{\psi}^{\rm pre}_{i'} \phi_{\rm post}^j \overline{\phi}_{\rm post}^{j'}
\end{equation}
or, simply, $\eta = \eta^\uparrow \otimes \eta_\downarrow$ with
\begin{equation}
    \eta^\uparrow= \ket{\psi^{\rm pre}}\bra{\psi^{\rm pre}}, \quad 
    \eta^\downarrow = \ket{\phi_{\rm post}}\bra{\phi_{\rm post}}.
\end{equation}
This situation corresponds to a system described by a two-state vector 
\begin{equation} \label{eq:two-state-vector}
    (\ket{\psi^{\rm pre}}, \bra{\phi_{\rm post}}),
\end{equation}
where $\ket{\psi^{\rm pre}}$ and $\bra{\phi_{\rm post}}$ evolve forward and backward in time, correspondingly.
This is the situation extensively studied in the framework of weak values and weak measurements~\cite{watanabe1955symmetry, aharonov1964time,aharonov1988result,aharonov2008two}.

A similar case of a mixed two-state vector~\cite{vaidman2017weak} 
$(\rho^{\rm pre}, \rho_{\rm post})$ corresponds to $\eta = \rho^{\rm pre}\otimes \rho_{\rm post}$, and can be realized physically either in an ancilla-free way by preparing $\rho^{\rm pre}$ on Q and then postselecting with $E_{\rm post}\propto \rho_{\rm post}$, or by employing several purifying ancillas as shown in Ref.~\cite{vaidman2017weak}.

We can also consider, so-called, generalized two-state vector~\cite{aharonov1991complete}, \BLUE{denoted as
\begin{equation}
    c_i^{j} (\ket{i}, \bra{j}),
\end{equation}
that describes a situation where} Q and A are initially prepared in a pure entangled state
\begin{equation}
    \ket{\Psi}_{\rm QA} = c_i^{j}\ket{i}_{\rm Q} \otimes \ket{j}_{A}
\end{equation}
($\rho^{\rm pre}=\ket{\Psi}_{\rm QA}\bra{\Psi}$)
and postselection corresponds to obtaining the maximally entangled Bell state
$E_{\rm post} \propto \ket{i}_{\rm Q}\bra{j} \otimes  \ket{i}_{\rm A}\bra{j}$.
The elements of $\eta$ then are given by
\begin{equation} \label{eq:elements-for-twostatevector}
    \eta_{ii'}^{jj'} = c_i^j\overline{c}_{i'}^{j'}
\end{equation}
or one can write 
\begin{equation}
    \eta = \ket{\Psi}\bra{\Psi}.
\end{equation}
Note that in this case forward and backward evolving parts appear to be entangled.

Next we can consider a randomized preparation and postselection scenario, studied in Ref.~\cite{silva2014pre}, where Alice randomly chooses to perform a postselection experiment with respect to generalized two-state vector 
$c^{(r)j}_i (\ket{i}, \bra{j})$
with probability $p^{(r)}$ ($p^{(r)}\geq 0$, $\sum_r p^{(r)}=1$).
This scenario corresponds to the two-state tensor in the form of a density vector:
\begin{equation} \label{eq:density-vector}
    \eta = \sum_r p^{(r)} \ket{\Psi^{(r)}}\bra{\Psi^{(r)}},
\end{equation}
with $\ket{\Psi^{(r)}} = c_i^{(r)j}\ket{i} \otimes \ket{j}$.
The obtained form of two-tensors is of particular importance, since, as we show further, any time-bidirectional state $\eta$ can be represented in the form of the density vector~\eqref{eq:density-vector}.

\subsection{Spectral decomposition and purification of time-bidirectional states}

Here we consider certain mathematical properties of time-bidirectional states and see how they are equivalent to bipartite states.
First, let us introduce multi-indices $\alpha:=(i,j)$, $\beta:=(i',j')$ and consider a time-bidirectional state $\eta_{ii'}^{jj'}$ as a matrix $\eta_{\alpha\beta}$.
Then one can see that according to the construction of $\eta$, for an arbitrary vector $\phi_{\gamma}$ of an appropriate dimension we have
\begin{equation}
    \overline{\phi}_\alpha \eta_{\alpha\beta} \phi_\beta \geq 0.
\end{equation}
Thus, $\eta_{\alpha\beta}$ is positive semi-definite, and so we can obtain its spectral decomposition
\begin{equation}
    \eta_{\alpha\beta} = \sum_r\lambda^{(r)} {\phi}^{(r)}_{\alpha}\overline\phi^{(r)}_{\beta}, \quad \lambda^{(r)} \geq 0,
\end{equation}
where $\phi^{(r)}$ form a set of orthonormal vectors ($\overline{\phi}^{(r)}_\alpha \phi^{(r')}_\alpha = \delta^{rr'})$.
Provided normalization~\eqref{eq:normalization}, we also have $\sum_r\lambda^{(r)}=1$.
Here we can also see that the only way to obtain $\sum_r\lambda^{(r)}=0$ is to have $\eta=0$.

After splitting multi-indices $\alpha$ and $\beta$ back, we come to
\begin{equation}~\label{eq:spectral-dec-for-eta}
    \eta_{ii'}^{jj'}=\sum_r\lambda^{(r)} \phi^{(r)j}_{i}\overline{\phi}^{(r)j'}_{i'},
\end{equation}
which is a two-state density vector [see also Eq.~\eqref{eq:density-vector}] introduced in Ref.~\cite{silva2014pre}.
We note that, similarly to spectral decomposition of mixed states, decomposition~\eqref{eq:spectral-dec-for-eta} completely determines measurement outcome probabilities for arbitrary measurements on $\eta$, yet does not specify exactly how $\eta$ is physically prepared.
Like an infinite number of possible statistical ensembles can yield the same mixed density matrix, `mixed' $\eta$ can be realized in an infinite number of possible ways.

Finally, we show that a time-bidirectional state $\eta$ also can `purified', i.e., obtained as a partial trace of some `pure' extended tensor $\widetilde{\eta}_{ii';rr'}^{jj'}$:
\begin{equation}
    \eta_{ii'}^{jj'} = \widetilde{\eta}_{ii';rr}^{jj'}, \quad
    \widetilde{\eta}_{ii';rr'}^{jj'}=\Theta_{i;r}^j \overline{\Theta}_{i';r'}^{j'}
\end{equation}
for some tensor $\Theta$.
This purifying tensor can be taken in the form 
\begin{equation}
   \Theta_{i;r}^j = \sqrt{\lambda^{(r)}} \phi_i^{(r)j}.
\end{equation}

From the physical point of view it can be realized as follows.
In the protocol, shown in Fig.~\ref{fig:ps_scheme}, Alice splits ancilla A into two particles: A and R, and prepare the pure state
\begin{equation}
    \ket{\Theta}_{\rm QAR}:=\Theta_{i;r}^j\ket{i}_{\rm Q}\ket{j}_{\rm A}\ket{r}_{\rm R}.
\end{equation}
Here R stands for a `reference' responsible for the purification. 
At the final step of the protocol, Alice performs the postselection with respect to Q and A in the maximally entangled state proportional to $\bra{i}_{\rm Q}\bra{i}_{\rm A}$, while keeping R untouched.
The effective state, observed by Bob, is then given by $\eta_{ii'}^{jj'} =\Theta_{i;r}^j \overline{\Theta}_{i';r}^{j'}$.

\section{Measuring observables} \label{sec:measurements}

In Sec.~\ref{sec:postsel-scheme}, we have considered a general scheme of an indirect measurement on Bob's side.
Here we focus on two particular cases of this measurement: The first one is a projective measurement and the second one is measurement of a weak value.

Before proceeding, we note some general properties of operation outcome tensor $K(\mu)$ and operation tensor $K$.
First, one can see that according to the construction of $K(\mu)$, we have
\begin{equation}
    \overline\phi_{i'}^{j'}K(\mu)^{ii'}_{jj}\phi_{i}^{j}\geq 0, 
\end{equation}
for arbitrary $\phi_{i}^{j}$.
This kind of semipositivity condition provides nonnegative probabilities for measurement outcome $\mu$.
The normalization condition for $K$ takes the form
\begin{equation}
    K^{ii'}_{kk} = \delta^{ii'}
\end{equation}
that actually is the standard normalization condition for Kraus operators. 
Note that $K^{kk}_{jj'}$ is not specified.
This asymmetry between upper and lower indices of $K$ catches an inherent time asymmetry of Bob's operations: In contrast to Alice's particle Q, Bob's particle P is in a preselected state and no postselection for P is considered.

\subsection{Projective measurements}

\begin{figure}
    \centering
    \includegraphics[width=\linewidth]{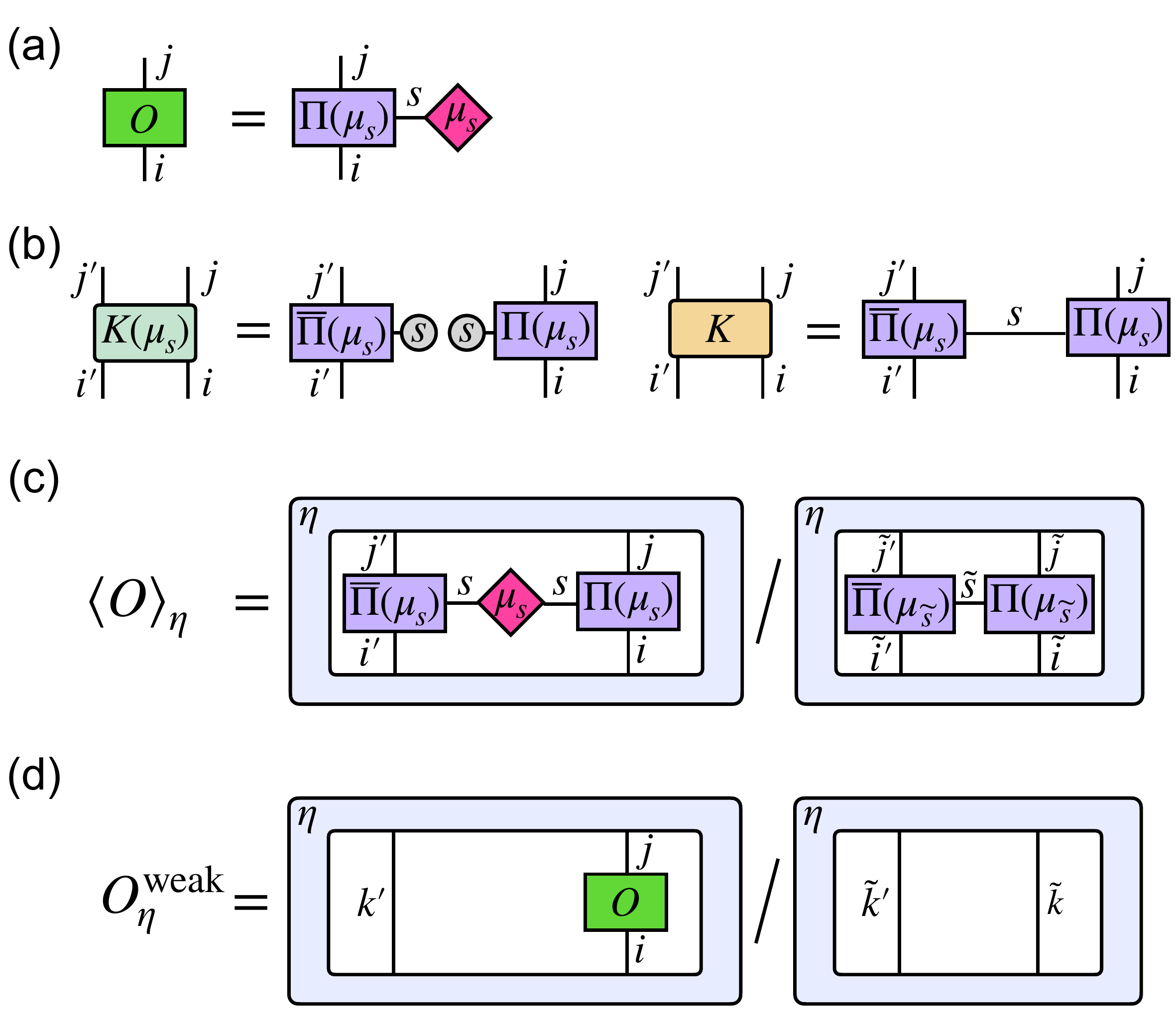}
    \caption{In (a) the spectral decomposition~\eqref{eq:spec-O} of the Hermitian observable $O$ is shown.
    In (b) an operation outcome $\mu_s$ and operation tensors are depicted.
    In (c) and (d) the expressions for mean~\eqref{eq:meanval} and weak~\eqref{eq:weak-gen} values of $O$ are depicted, correspondingly.}
    \label{fig:tensor-meas}
\end{figure}

Consider a finite-dimensional Hermitian observable $O$ with a spectral decomposition of the form:
\begin{equation} \label{eq:spec-O}
    O = \sum_s \mu_s \Pi(\mu_s),
\end{equation}
where $\{\mu_s\}$ is a set of distinct real numbers (the size of $\{\mu_s\}$ does not exceed the dimensionality of $O$), and $\Pi(\mu_s)=\Pi^{i}_j(\mu_s)\ket{i}\bra{j}$ are orthogonal projectors forming a complete set ($\sum_s \Pi(\mu_s) = \mathbb{1}$, $\Pi(\mu_{s})\Pi(\mu_{s'})=\delta_{ss'}\Pi(\mu_{s})$).
The corresponding tensor network diagram is presented in Fig.~\ref{fig:tensor-meas}(a).
According to the axiomatics of quantum mechanics, measuring $O$ of a pure state $\ket{\psi}$ provides a real value $\mu_i$ with probability $P(\mu_i)=\bra{\psi}\Pi(\mu_i)\ket{\psi}$.
At the same time, the original state $\ket{\psi}$ `collapses' to $P(\mu_i)^{-1/2}\;\Pi(\mu_i)\ket{\psi}$.

\BLUE{From the viewpoint of the scheme from Fig.~\ref{fig:ps_scheme}, this kind of measurement is equivalent to the scenario in which $\rho^{\rm probe}=\ket{0}_{\rm P}\bra{0}$, where $\ket{0}$ is some fixed state; $U$ makes an isometry transformation of the form
\begin{equation}
    \ket{0}_{\rm P}\otimes \ket{\psi}_{\rm Q} \mapsto
    \sum_s\ket{s}_{\rm P}\otimes \Pi(\mu_s)\ket{\psi}_{\rm Q},
\end{equation}
where $\ket{\psi}$ is an arbitrary state and $\{\ket{s}\}$ forms an orthonormal subset of states (transformation in the remaining space can be arbitrary satisfying unitarity of $U$), and POVM $E$ consists of effects $E(\mu_s)=\Pi(\mu_s)$.}

One can easily check that in this case, the outcome $\mu_i$ tensor and operation tensor are correspondingly given by
\begin{equation}
    K(\mu_i) = \Pi(\mu_i) \otimes \overline{\Pi}(\mu_i), \quad     
    K = \sum_s \Pi(\mu_s) \otimes \overline{\Pi}(\mu_s)
\end{equation}
[see also Fig.~\ref{fig:tensor-meas}(b)].

Then for an arbitrary time-bidirectional state $\eta$, the probability to obtain $\mu_i$ is given by $P(\mu_i)=(K\bullet \eta)^{-1} K(\mu_i)\bullet \eta$, and the mean value of $O$ takes the form
\begin{equation} \label{eq:meanval}
    \langle O \rangle_\eta =\sum_s P(\mu_s)\mu_s= \frac{\widetilde O \bullet \eta}{K \bullet \eta},
\end{equation}
where we have introduced an auxiliary projective measurement tensor 
\begin{equation} \label{eq:spec-O-ext}
    \widetilde O := \sum_s  \mu_s \Pi(\mu_s) \otimes \overline{\Pi}(\mu_s)
\end{equation}
[see also Fig.~\ref{fig:tensor-meas}(c)].

One can see that in the case of no postselection, time-bidirectional state $\eta = \rho^{\rm pre}_{\rm Q} \otimes \rho_{\rm mix}$, the expression for the mean value reduces to
\begin{equation}
    \langle O \rangle_\eta={\rm Tr}[O\rho^{\rm pre}_{\rm Q}].
\end{equation}
We also note the decomposition~\eqref{eq:spec-O-ext} can be considered as a generalization of a standard spectral decomposition~\eqref{eq:spec-O} to the time-bidirectional case.

\subsection{Weak values}

We call a weak value of a Hermitian observable $O=\sum_{ij}O^i_j\ket{i}\bra{j}$ with respect to a time-bidirectional state $\eta$ a quantity
\begin{equation} \label{eq:weak-gen}
    O^{\rm weak}_\eta :=
    \frac{\eta_{ik'}^{jk'}O^{i}_j}{
    \eta_{\tilde i \tilde k'}^{\tilde i \tilde k'}
    }=
    \frac{(\mathbb{1}\otimes O)\bullet\eta}{(\mathbb{1}\otimes \mathbb{1})\bullet\eta}
\end{equation}
[see also Fig.~\ref{fig:tensor-meas}(d)].
To justify this definition, we first note that for $\eta= \ket{\psi^{\rm pre}}\bra{\psi^{\rm pre}} \otimes \ket{\phi_{\rm post}}\bra{\phi_{\rm post}}$, which corresponds to a standard two-state vector~\eqref{eq:two-state-vector}, Eq.~\eqref{eq:weak-gen} reduces to
\begin{equation}
    O^{\rm weak}_\eta = \frac{\psi^{\rm pre}_{i}\overline{\psi}^{\rm pre}_{k} \phi_{\rm post}^j \overline{\phi}_{\rm post}^{k}O^{i}_j}{\psi^{\rm pre}_{\tilde i}\overline{\psi}^{\rm pre}_{\tilde k} \phi_{\rm post}^{\tilde i} \overline{\phi}_{\rm post}^{\tilde k}} = 
    \frac{\bra{\phi_{\rm post}}O\ket{\psi^{\rm pre}}}{\braket{\phi_{\rm post}|{\psi^{\rm pre}}}},
\end{equation}
which is the standard definition of the weak value for the two-state vector $(\ket{\psi^{\rm pre}},\bra{\phi_{\rm post})}$~\cite{aharonov1988result,aharonov2008two}.
Then we show that standard experimental setups devoted to extracting real and imaginary parts of a weak value~\cite{dressel2014colloquium} provide real and imaginary parts of $O^{\rm weak}_\eta$ for general $\eta$ correspondingly.

Let P in our postselection experiment now be a one dimensional particle and let ${\cal Q}$ and ${\cal P}$ be its coordinate and momentum operators correspondingly.
We consider P initially prepared in a pure state $\rho^{\rm probe}=\ket{\psi}\bra{\psi}$ with a Gaussian wave function
\begin{equation}
    \braket{q|\psi} = \frac{1}{(2\pi\sigma)^{1/4}}\exp\left(-\frac{q^2}{4\sigma^2}\right),
\end{equation}
where $\ket{q}$ is an eigenstate of ${\cal Q}$ with an eigenvalue $q$, and $\sigma^2>0$. 
Let the coupling unitary operation $U$ be realized by applying a measurement Hamiltonian ${\cal H}={\cal P}\otimes O$ during some small time $\epsilon \ll 1$ (here we employ dimensionless units with $\hbar\equiv 1$).
Considering orders of $\epsilon$ not higher than the first, we have
\begin{equation}
    U=e^{-\imath \epsilon  O\otimes\cal{P}}
    \approx \mathbb{1}-\imath \epsilon O\otimes\cal{P}
\end{equation}

The mean value of coordinate $q$, given passing of the postselection, takes the form
\begin{multline}
    \langle {\cal Q} \rangle_{\eta} = \frac{(\mathbb{1}\otimes\mathbb{1})\bullet \eta \braket{{\cal Q}}_\psi}{
    [\mathbb{1}\otimes\mathbb{1}-
    \imath\epsilon (\mathbb{1}\otimes O \braket{\cal{P}}_\psi-\overline O \otimes \mathbb{1} \braket{{\cal P}}_\psi)]\bullet \eta
    }\\
    -\imath\epsilon
    \frac{(\mathbb{1}\otimes O \braket{{\cal Q}\cal{P}}_\psi- \overline O \otimes \mathbb{1} \braket{{\cal P}\cal{Q}}_\psi))\bullet \eta}{
    [\mathbb{1}\otimes\mathbb{1}-
    \imath\epsilon (\mathbb{1}\otimes O \braket{\cal{P}}_\psi-\overline O \otimes \mathbb{1} \braket{{\cal P}}_\psi]\bullet \eta
    },
\end{multline}
where $\braket{\cdot}_\psi\equiv \bra{\psi}\cdot\ket{\psi}$.
Taking into account that $\braket{{\cal P}}_\psi=\braket{O}_\psi=0$, $\braket{{\cal QP}}_\psi=-\braket{{\cal PQ}}_\psi=\imath/2$, and $(\mathbb{1}\otimes O)\bullet \eta = \overline {(\overline{O}\otimes\mathbb{1})\bullet \eta}$,
we come to 
\begin{equation}
    \langle {\cal Q} \rangle_{\eta} = \epsilon{\rm Re}O^{\rm weak}_\eta.
\end{equation}
In a similar way, provided $\braket{{\cal P}^2}_\psi=1/(4\sigma^2)$, measuring the (conditional) mean value of momentum gives
\begin{equation}
    \langle {\cal P}\rangle_\eta  = \frac{\epsilon}{4\sigma^2} {\rm Im}O^{\rm weak}_\eta.
\end{equation}
We then see that experimentally-accessible real values of $\langle {\cal Q}\rangle_\eta$ and $\langle {\cal P}\rangle_\eta$, provided definition~\eqref{eq:weak-gen}, allow reconstructing complex value of $O^{\rm weak}_\eta$ for the general time-bidirectional state $\eta$.

\section{Tomography of a single-qubit time-bidirectional state} \label{sec:tomography}

Here we consider the problem of an experimental reconstruction of an unknown time-bidirectional state.
Similarly to the case of a standard quantum state tomography, the reconstruction of a time-bidirectional state $\eta$ requires its measuring with some sufficient informationally complete collection of outcome tensors sets $\{{\bf K}^{(r)}\}_r$, where ${\bf K}^{(r)}=\{K^{(r)}(\mu')\}_{\mu'}$ denotes a particular set of outcome tensors.
The information completeness condition ensures that observed probabilities
\begin{equation} \label{eq:prob-for-tom}
    P^{(r)}(\mu) = \frac{K^{(r)}(\mu) \bullet \eta}{K^{(r)} \bullet \eta}
\end{equation}
are sufficient for recovering all elements of $\eta$.
An unknown time-bidirectional state can be recovered, e.g., by maximizing likelihood function
\begin{equation}
    {\cal L}(\eta) =  \sum_{r,k}\mu^{(r,k)}\log P^{(r)}(\mu^{(r,k)}),
\end{equation}
where $\{\mu^{(r,k)}\}_k$ denotes a particular measurement outcome obtained in ${\bf K}^{(r)}$ measurement.
We note this problem can be efficiently solved with gradient descent within a corresponding Riemannian manifold of positive semi-definite matrices~\cite{pechen2008control,oza2009optimization,luchnikov2021qgopt, luchnikov2021riemannian}. 

Here we focus on constructing practical schemes for the time-bidirectional state tomography of a single qubit (the general, yet less detailed, approach for tomography of arbitrary finite-dimensional time-bidirectional states can be found in Ref.~\cite{silva2014pre}).
We consider two basic techniques borrowed from the standard quantum state tomography: The first one is based on MUB and employs several measurement configurations, and the second one is based on SIC-POVM measurements and realized with a single set of operation outcome tensors.
We note that the presented schemes can be directly generalized to multiqubit time-bidirectional states.

Before proceeding further, we highlight some aspects of the time-bidirectional states tomography that are different compared to the standard quantum state tomography.
First, the standard Born rule is replaced with Eq.~\eqref{eq:prob-for-tom}.
Second, since we actually deal with two states propagating in opposite time directions, the performed measurements have to be nondestructive.
A convenient way to achieve this property is to couple the measured particle to some ancillary particle (like P in Fig.~\ref{fig:ps_scheme}), and then perform the final read-out measurement on this ancilla. 
Third, the measurement employed in the tomography generally affects the postselection probability.
It may turn out that that for some measurements and some time-bidirectional states, the postselection probability can be zero.
This fact complicates an analysis of statistical size effects, which we leave for future study.

\subsection{MUB-based approach}

In the case of a single qubit, MUB correspond to eigenvectors of standard Pauli operators
\begin{equation}
    \sigma_x=\begin{bmatrix} 0 & 1 \\ 1 & 0  \end{bmatrix}, \quad 
    \sigma_y=\begin{bmatrix} 0 & -\imath \\ \imath & 0  \end{bmatrix}, \\
    \quad
    \sigma_z=\begin{bmatrix} 1 & 0 \\ 0 & -1  \end{bmatrix}.
\end{equation}
The corresponding projectors on eigenspaces read
\begin{equation}
    \Pi^{(r)}(\mu)=\frac{1}{2}(\mathbb{1}+\mu\sigma_r), \quad \mu=\pm1, \quad r=x,y,z
\end{equation}
and provide spectral decompositions of the form
\begin{equation}
    \sigma_r=\Pi^{(r)}(1)-\Pi^{(r)}(-1).
\end{equation}

As already noted, in contrast to standard single qubit quantum state tomography, where an unknown qubit state is measured in three MUB (corresponding to $x$-,$y$-, and $z$- projection on the Bloch sphere), in the case of time-bidirectional states we actually deal with states of two qubits propagating in opposite time directions.
Thereby, we measure forward and backward propagating states in two different MUB.
To do so, we take operation outcome tensors of the form
\begin{multline}~\label{eq:mub_operator_full}
    K^{(r_1,r_2)}(\boldsymbol\mu)=\Pi^{(r_2)}(\mu_2)\Pi^{(r_1)}(\mu_1) \\ \otimes 
    \overline\Pi^{(r_2)}(\mu_2)\overline\Pi^{(r_1)}(\mu_1)
\end{multline}
with $\boldsymbol \mu = (\mu_1, \mu_2)$, where $\mu_1$ and $\mu_2$ correspond to measurement results of a forward and backward evolving parts respectively.

If $r_1\neq r_2$, Eq.~\eqref{eq:mub_operator_full} transform to
\begin{multline}
    K^{(r_1,r_2)}(\boldsymbol \mu) = \frac{1}{2} \ket{\psi^{(r_2)}(\mu_2)}\bra{\psi^{(r_1)}(\mu_1)} \\
    \otimes 
    \ket{\overline\psi^{(r_2)}(\mu_2)}\bra{\overline\psi^{(r_1)}(\mu_1)},
\end{multline}
where $\ket{\psi^{(r)}(\mu)}$ denotes an eigenvector of $\sigma_r$ with eigenvalue $\mu$.
The operation tensor in this case is given by
\begin{equation}
    K^{(r_1,r_2)ii'}_{jj'} = \sum_{\mu_1 \mu_2}K^{(r_1,r_2)}(\mu_1,\mu_2)^{ii'}_{jj'}=\frac{1}{2} \delta^{ii'} \delta_{jj'}.
\end{equation}
Note that it provides nonzero postselection probability for every $\eta\neq 0$.

For $r_1=r_2$, we have
\begin{multline}  \label{eq:same-r_i}
    K^{(r_1,r_2)}(\boldsymbol \mu) = \delta_{\mu_1,\mu_2} \ket{\psi^{(r_1)}(\mu_1)}\bra{\psi^{(r_1)}(\mu_1)}\\
    \otimes 
    \ket{\overline{\psi}^{(r_1)}(\mu_1)}\bra{\overline{\psi}^{(r_1)}(\mu_1)},
\end{multline}
i.e. this measurement corresponds to the single projective measurement of $\sigma_{r_1}$.
The corresponding operation tensor takes the form:
\begin{equation}
    K^{(r_1,r_2)} = \sum_{\mu=\pm1} \Pi^{(r_1)}(\mu)\otimes \overline{\Pi}^{(r_1)}(\mu).
\end{equation}
Note that for some time-bidirectional states, the postselection probability for this kind of measurements turns into zero (i.e., $K^{(z,z)}\bullet \eta=0$ for $\eta=\ket{0}\bra{0}\otimes \ket{1}\bra{1}$).

\begin{figure*}
    \centering
    \includegraphics[width=0.9\linewidth]{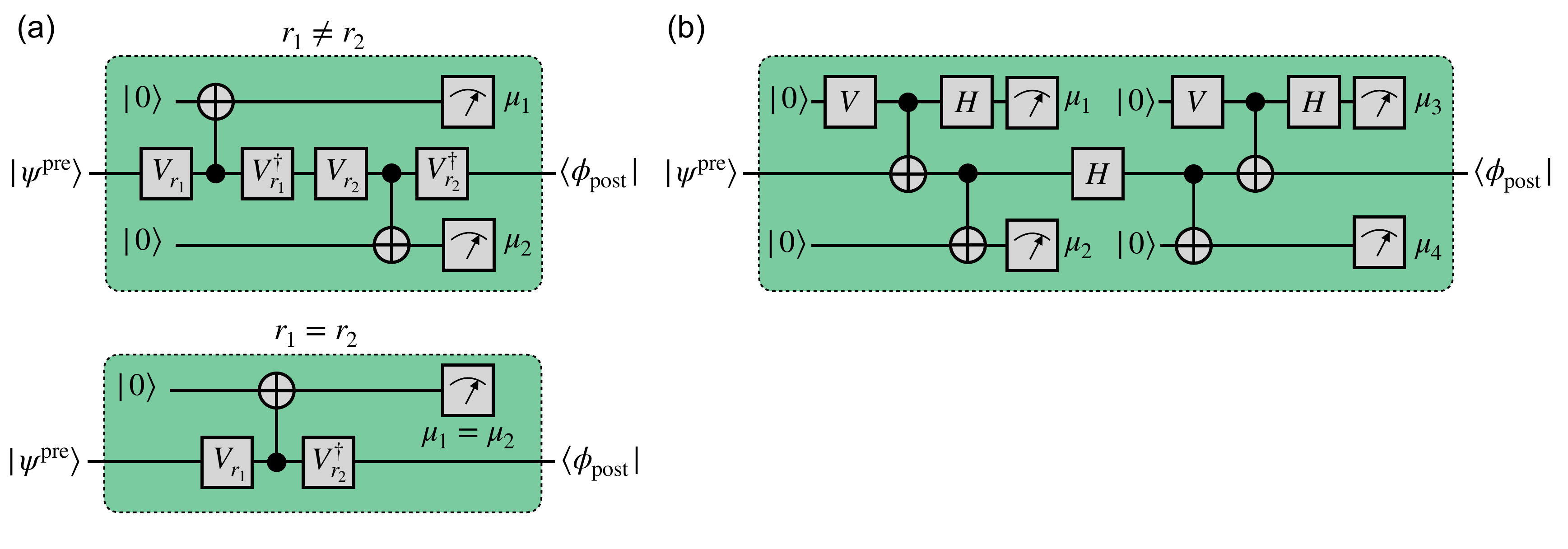}
    \caption{The circuit representation of the MUB-based (a) and the SIC-POVM-based (b) tomography schemes of a single-qubit time-bidirectional state.
    Forward and backward evolving parts of the time-bidirectional state are labeled by $\ket{\psi^{\rm pre}}$ and $\ket{\phi_{\rm post}}$, correspondingly.
    Standard notations for controlled-NOT gate, Hadamard gate, and computational basis measurements are used; $V_{r_i}$ and $V$ are given by Eq.~\eqref{eq:Vri} and Eq.~\eqref{eq:V}, respectively.}
    \label{fig:tomography-circuits}
\end{figure*}

The described two-MUB measurement can be realized via quantum circuits shown in Fig.~\ref{fig:tomography-circuits}(a).
Here we employ two ancillary qubits, initialized in $\ket{0}$, in order to realize nondestructive projectile Pauli measurements.
The coupling is performed with controlled-NOT (CNOT) gates, surrounded by `basis change' operators $V_{r_i}$.
Choosing these unitaries in the form
\begin{equation} \label{eq:Vri}
    V_x := R_y(\pi/2), \quad, V_y := R_x(\pi/2) \quad V_z := \mathbb{1},
\end{equation}
where $R_r(\theta)=e^{-\imath \sigma_r \theta/2}$ denotes a standard rotation operation, we obtain proper $x$-, $y$-, and $z$-Pauli measurements respectively (computational basis measurement outcomes 0 and 1 have to interpreted as $+1$ as $-1$ correspondingly).
Note that due to Eq.~\eqref{eq:same-r_i}, in the case of $r_1=r_2$, only a single ancilla is required.

In total, one needs nine circuits, corresponding to all combinations of $r_1$ and $r_2$, to recover an unknown single-qubit time-bidirectional state. 
Next, we will see how to cope with the same problem with a single circuit.

\subsection{SIC-POVM-based approach}

\begin{figure}
    \centering
    \includegraphics[width=0.6\linewidth]{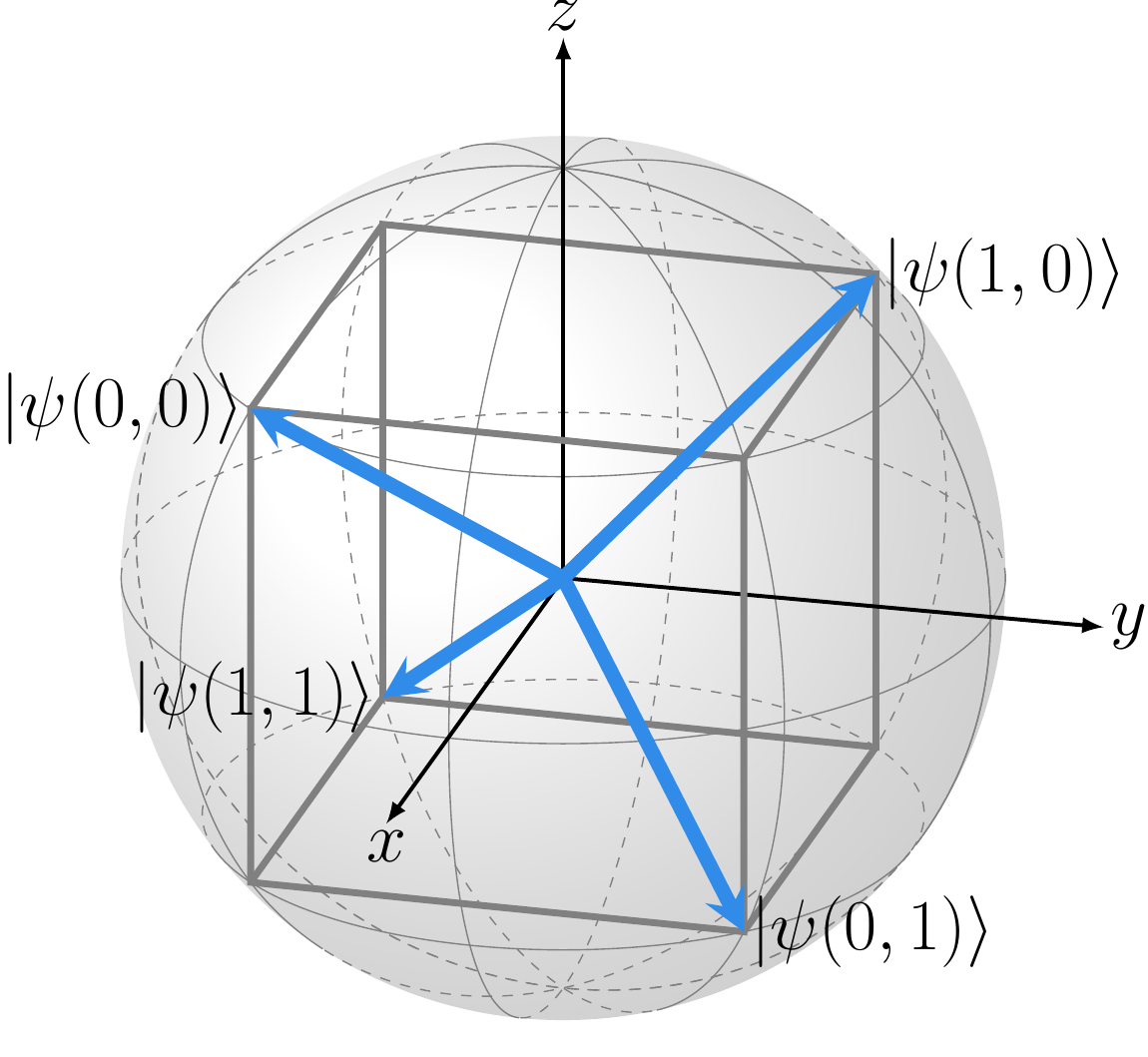}
    \caption{Pure states used for constructing SIC-POVM elements.}
    \label{fig:sicpovm_bloch}
\end{figure}

An alternative approach for recovering an unknown state is to make a single informationally complete measurement, e.g. given by a SIC-POVM.
In the case of a single qubit, SIC-POVM is defined by four pure states corresponding to vertices of a regular tetrahedron inscribed in the Bloch sphere.
For our purpose, we take these states in the following form:
\begin{equation}
    \ket{\psi(\mu_1,\mu_2)}=\sigma_x^{\mu_2}\sigma_z^{\mu_1}(\cos\frac{\theta}{2}\ket{0}+e^{-\imath\pi/4}\sin\frac{\theta}{2}\ket{1}),
\end{equation}
where $\mu_i\in\{0,1\}$ and $\theta:=\arccos(1/\sqrt{3})$
(see also Fig.~\ref{fig:sicpovm_bloch}).
The corresponding SIC-POVM is given by a set of four operators $\{\Pi(\mu_1,\mu_2)/2\}_{\mu_1,\mu_2}$ with
\begin{equation}
    \Pi(\mu_1,\mu_2)=\ket{\psi(\mu_1,\mu_2)}\bra{\psi(\mu_1,\mu_2)}.
\end{equation}

In order to recover both forward and backward evolving states, we perform a doubled SIC-POVM measurement with outcome tensors of the form 
\begin{multline}
 \label{eq:SIC-POVM-outcome-tensor}
	K(\boldsymbol\mu) = \frac{1}{8}
	\ket{\psi(\mu_1,\mu_2)}\bra{\overline\psi(\mu_3,\mu_4)} \\ \otimes \ket{\overline\psi(\mu_1,\mu_2)}\bra{\psi(\mu_3,\mu_4)}
\end{multline}
where $\boldsymbol\mu=(\mu_1,\mu_2,\mu_3,\mu_4)$ and $\mu_i\in\{0,1\}$.
The corresponding operation tensor reads
\begin{equation}
    K^{ii'}_{jj'}=
    \sum_{\boldsymbol\mu} K(\boldsymbol\mu)^{ii'}_{jj'} = \frac{1}{2} \delta^{ii'} \delta_{jj'}
\end{equation}
and provides a nonzero postselection probability for every nontrivial time-bidirectional state $\eta \neq 0$.
Note that provided normalization rule~\eqref{eq:normalization}, $K \bullet \eta$ is the same for every $\eta$, which makes probability~\eqref{eq:prob-for-tom} of the same form as the Born rule for standard two-qubit states.

The circuit implementing the considered `doubled' SIC-POVM measurement in shown in Fig.~\ref{fig:tomography-circuits}(b).
To construct this circuit, the scheme for a standard SIC-POVM measurement from Ref.~\cite{kiktenko2020probability} is used.
Here, single-qubit unitary $V$ makes the transformation
\begin{equation} \label{eq:V}
	V\ket{0}=\cos\frac{\theta}{2}\ket{0}+e^{\imath\pi/4}\sin\frac{\theta}{2}\ket{0}
\end{equation}
and can be taken in the form $V=R_z(\pi/4)R_y(\theta)$. 
To make this measurement, one needs at most four ancillary qubits for storing all elements $\boldsymbol\mu$.
We note that the number of ancillas can be reduced by using qubits reinitialization.
In Fig.~\ref{fig:tomography-circuits}(b) we show the scheme with two ancillas, reinitializing them back into $\ket{0}$ after reading out the values of $\mu_1$ and $\mu_2$.

\section{Tracking time-reversal state propagation in a quantum teleportation protocol} \label{sec:experiment}

Here we demonstrate how the developed formalism, as well as developed tomography techniques, allows one to observe a time-reversal propagation of quantum states.
Consider a standard single-qubit quantum teleportation protocol~\cite{bennett1993teleporting}.
Let Alice be given some pure qubit state $\ket{\psi}$ on particle ${\cal A}$.
Let also Alice and Bob share maximally entangled state $\ket{\Phi^+}$ on particles ${\cal B}$ and ${\cal C}$, s.t. ${\cal B}$ goes to Alice, and ${\cal C}$ belongs to Bob.
Here and hereafter, standard notation for Bell states is used:
\begin{equation}
	\ket{\Phi^{\pm}} \equiv \frac{1}{\sqrt{2}}(\ket{00}\pm\ket{11}), \quad \ket{\Psi^{\pm}} \equiv \frac{1}{\sqrt{2}}(\ket{01}\pm\ket{10}).
\end{equation}
In order to transmit $\ket{\psi}$ to Bob, Alice measures her particles ${\cal A}$ and ${\cal B}$ in the Bell basis and transmits to Bob the obtained outcome encoded in two bits via a classical channel.
By applying a proper single qubit unitary on ${\cal C}$, which depends on the message from Alice, Bob obtains ${\cal C}$  exactly in the state $\ket{\psi}$.

The natural question is how the state $\ket{\psi}$ propagates from ${\cal A}$ to ${\cal C}$.
Obviously, it can not propagate with the classical message from Alice, which is uncorrelated with $\ket{\psi}$: Four outcomes of Bell measurement appear with the same probability of 1/4.
The only quantum medium that connects Alice and Bob is the entangled pair of ${\cal B}$ and ${\cal C}$, so one can suggest that $\ket{\psi}$ propagates from ${\cal A}$ to ${\cal C}$ via ${\cal B}$.

\begin{figure*}
    \centering
    \includegraphics[width=\linewidth]{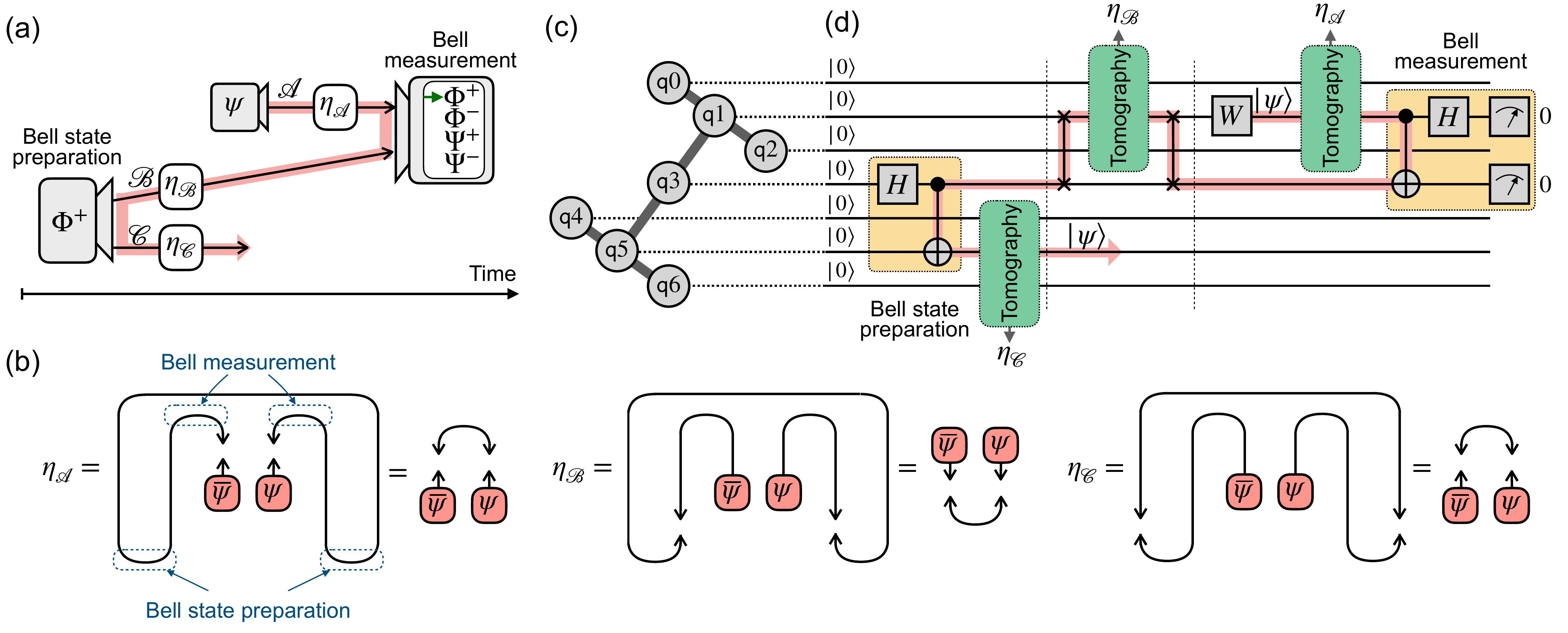}
    \caption{In (a) the scheme of postselective quantum teleportation protocol is depicted.
    If the postselection is performed with respect to Bell measurement outcome $\ket{\Phi^+}$ on ${\cal A}$ and ${\cal B}$, the teleported state $\ket{\psi}$ can appear on ${\cal C}$ even before its preparation by on ${\cal A}$. 
    In (b) time-bidirectional states $\eta_{\cal A}$, $\eta_{\cal B}$, and $\eta_{\cal C}$ are depicted.
    No possible decoherence effects are taken into account.
    In (c) the architecture of seven-qubit superconducting processor {\sf ibm\_oslo} is shown. 
    Connections between qubits ${\sf q}0, \ldots, {\sf q}6$ correspond to an ability to apply CNOT gates.
    In (d) circuits for the tomography of $\eta_{\cal A}$, $\eta_{\cal B}$, and $\eta_{\cal C}$ in the teleportation protocol, shown in (a), are depicted.
    Vertical dashed lines shows `barriers', which ensure that each next part of the circuit is realized after the previous one.
    The path of the state $\ket{\psi}$ through the circuit is highlighted.
    }
    \label{fig:teleporation}
\end{figure*}

A hint for this problem can be obtained by considering a fixed outcome of Alice's measurement.
It is especially helpful to consider the case, where Alice obtains a particular outcome $\ket{\Phi^+}$.
In this case, the corresponding Bob's single-qubit transformation is equal to the identity.
It means that Bob already has his particle ${\cal C}$ in the state $\ket{\psi}$.
Since no transformations were applied to ${\cal C}$, ${\cal C}$ appears to be in the state $\ket{\psi}$ just after the birth of the Bell pair.
Note that this remains to be true even if the state $\ket{\psi}$ is prepared on ${\cal A}$ after the Bell pair's birth as is shown in Fig.~\ref{fig:teleporation}(a).
In this case we effectively have a travel of $\ket{\psi}$ from ${\cal A}$ back in time to the moment of Bell pair birth  via ${\cal B}$, and then forward in time on ${\cal C}$.

In order to justify this conclusion, let us consider time-bidirectional states $\eta_{\cal A}$, $\eta_{\cal B}$, and $\eta_{\cal C}$ of the corresponding particles $\cal A$, $\cal B$, and $\cal C$, given the postselection condition of measuring $\ket{\Phi^+}$ in Alice's Bell measurement [see Fig.~\ref{fig:teleporation}(a)].
Notably, we define $\eta_{\cal B}$ and $\eta_{\cal C}$ before the preparation of ${\cal A}$ in $\ket{\psi}$ from the viewpoint of an external macroscopic observer.

One can easily check that the introduced time-bidirectional states read
\begin{equation} \label{eq:etaABC}
	\eta_{\cal A} = \eta_{\cal C}  = \ket{\psi}\bra{\psi} \otimes \rho_{\rm mix} \quad \eta_{\cal B} = \rho_{\rm mix} \otimes \ket{\psi}\bra{\psi}
\end{equation}
[see also Fig.~\ref{fig:teleporation}(b)].
The form of these tensors justifies our suggestion about the time travel of $\ket{\psi}$ via the route ${\cal A}\rightarrow {\cal B} \rightarrow {\cal C}$.
Note that $\ket{\psi}\bra{\psi}$ is in the forward evolving part of $\eta_{\cal A}$ and $\eta_{\cal C}$, and is in the backward evolving part of $\eta_{\cal B}$.

To verify our consideration, we perform a demonstration on the superconducting seven-qubit processor {\sf ibm\_oslo} provided with cloud access by IBM.
\BLUE{In this way, each time-bidirectional state $\eta_{\cal A}$, $\eta_{\cal B}$, and $\eta_{\cal C}$ describes two copies of a two-dimensional subspace spanned by the ground state and the first excited state of superconducting anharmonic oscillators (transmon qubits).}
The architecture of the processor is shown in Fig.~\ref{fig:teleporation}(c).
Here the connections between physical qubits, denoted by ${\sf q}0, \ldots, {\sf q}6$, correspond to an ability to apply CNOT gates. 
The circuits realizing tomography of qubits in a quantum teleportation protocol are shown in Fig.~\ref{fig:teleporation}(d).
We use pairs ${\sf q}0, {\sf q}2$ and  ${\sf q}4, {\sf q}6$ in order to make SIC-POVM-based tomography according to Fig.~\ref{fig:tomography-circuits}(b), while ${\sf q}1$, ${\sf q}3$, and ${\sf q}5$ are used for storing states of ${\cal A}$, ${\cal B}$, and ${\cal C}$.
Initially the Bell pair of ${\cal B}$ and ${\cal C}$ is prepared on ${\sf q}3$ and ${\sf q}5$ correspondingly.
Then in order to make tomography of ${\cal B}$, we move it from ${\sf q}3$ to ${\sf q}1$ by applying the SWAP gate and then return it back on ${\sf q}3$ by another SWAP gate.
Next, ${\cal A}$ is prepared on ${\sf q}1$ in the state $\ket{\psi}=\begin{bmatrix} \sqrt{3}/2 & {e^{\imath \pi/4}}/{2} \end{bmatrix}^{\rm T}$
%\begin{equation}
%    \ket{\psi}=\frac{\sqrt{3}}{2}\ket{0}+\frac{e^{\imath \pi/4}}{2}\ket{1}
%\end{equation} 
by applying $W=R_z(\pi/4) R_x(\pi/2) R_z(\pi/3) R_x(\pi/2)$ to $\ket{0}$.
Finally, the Bell measurement on ${\sf q}1$ and  ${\sf q}3$ is performed.
The postselection is made with respect to the 0 for both outcomes, which corresponds to $\ket{\Phi^+}$ outcome.
We apply barriers [denoted by vertical dashed lines in Fig.~\ref{fig:teleporation}(d)] to ensure that the tomography measurements of $\eta_{\cal B}$ and ${\cal C}$ are performed before preparation of $\ket{\psi}$.
Also note that in the case of $\eta_{\cal A}$ and $\eta_{\cal C}$ tomography, an additional barrier between SWAP gates is put in order to prevent their removing by a transpiler.
In total, three circuits corresponding to the tomography of $\eta_{\cal A}$, $\eta_{\cal B}$,  and $\eta_{\cal C}$, are run.
For each circuit, $N=20000$ shots (numbers of run) are used.
Transpiled versions of three launched circuits, obtained measurement counts, and calibration data are also available at~\cite{supplementary}.

\begin{figure*}
    \centering
    \includegraphics[width=\linewidth]{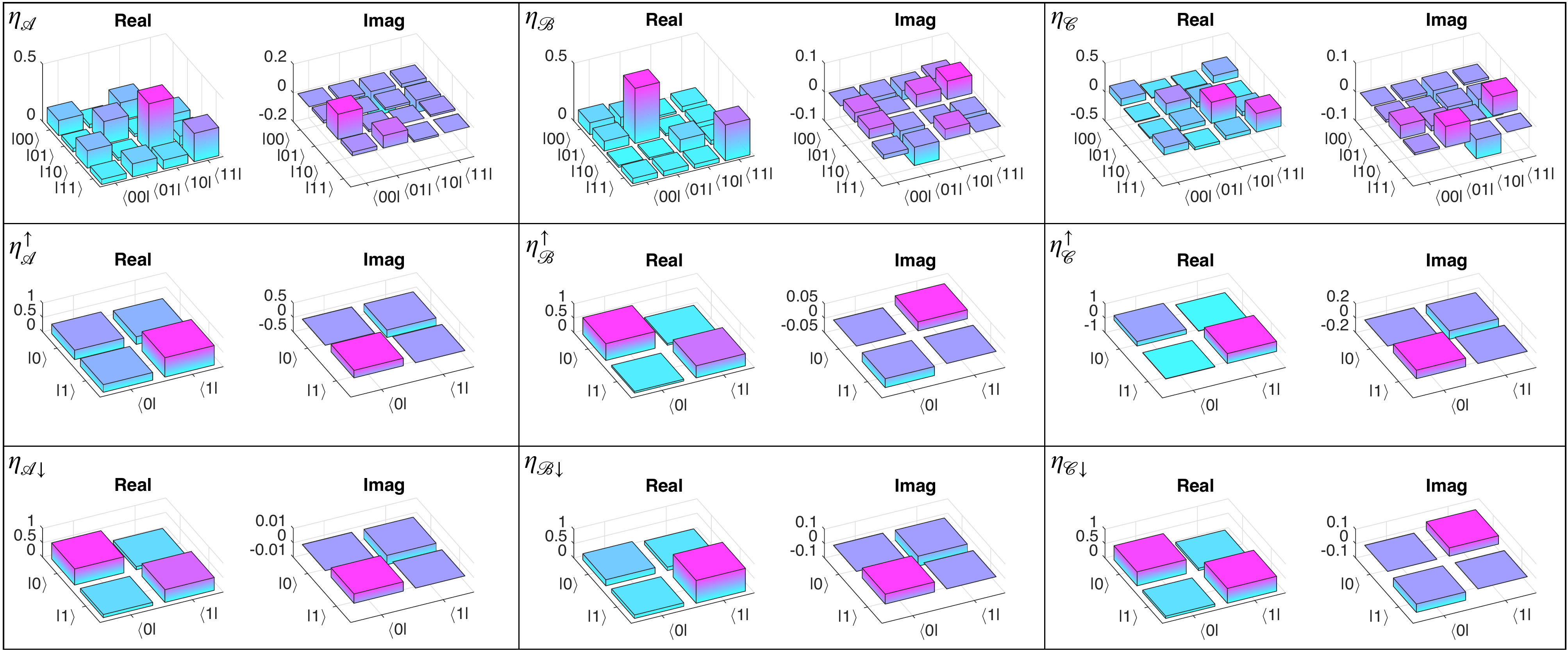}
    \caption{The obtained results of time-bidirectional states tomography. 
    Forward and backward evolving single-qubit parts are obtained by corresponding contraction of the reconstructed time-bidirectional states.
    %The color of each bar corresponds to its height, and is used just to emphize an 
    }
    \label{fig:tomography}
\end{figure*}

The tomography results are shown in Fig.~\ref{fig:tomography}.
Here we show both the full time-bidirectional states and their reduced forward and backward evolving parts.
One can see that obtained tensors are noisy versions of tensors given by Eq.~\eqref{eq:etaABC}.
This fact can be explained by an influence of decoherence processes and effects of the finite length statistics.
We note that decoherence affects both the quality of gates implementation, especially two-qubit ones, and read-out measurements. 
It affects a `purity' of the postselection, and accuracy of the tomography.
We also note that approximately $N/4$ outcomes are used in the tomography due to the postselection condition.

The corresponding fidelities with respect to ideal states, and linear entropies, defined as $S(\rho)=1-{\rm Tr}(\rho^2)$), are shown in Table~\ref{tab:rslts}.
One can see that the fidelity of the state $\ket{\psi}$ drops down during its travel through the time-reversal path:
It starts at 0.91 on  $\eta_{\cal A}^{\uparrow}$, then is reduced to 0.76 on $\eta_{\cal B\downarrow}$, and finally drops to 0.64 on $\eta_{\cal C}^\uparrow$.
At the same time, the linear entropy increases from 0.16 through 0.29 to 0.40.
This state corruption corresponds to a propagation of $\ket{\psi}$ along its own `thermodynamic' time arrow different from the time arrow of a classical observer.
So we see that the TBSF allows one to study irreversible processes occurring with quantum states during their travel along nontrivial postselection-induced space-time trajectories.

\begin{table}
    \centering
    \begin{tabular}{c|c|c}
         State & Fidelity & Linear entropy  \\ \hline\hline
         $\eta_A$ & 0.89 & 0.53 \\
         $\eta_A^{\uparrow}$ & {\bf 0.91} & {\bf 0.16} \\
         $\eta_{A\downarrow}$ & 0.98 & 0.47 \\\hline
         $\eta_B$ & 0.74 & 0.61 \\
         $\eta_B^{\uparrow}$ & 0.99 & 0.48 \\
         $\eta_{B\downarrow}$ & {\bf 0.76} & {\bf 0.29} \\\hline
         $\eta_C$ & 0.62 & 0.63 \\
         $\eta_C^{\uparrow}$ & {\bf 0.64} & {\bf 0.40}\\
         $\eta_{C\downarrow}$ & 0.99 & 0.48 \\
    \end{tabular}
    \caption{Properties of the reconstructed states.
    The results for reduced tensors carrying $\ket{\psi}$ are highlighted.
    }
    \label{tab:rslts}
\end{table}

We note that the observed behavior can be modeled, e.g., by adding depolarizing noise to the Bell state preparation and measurement.
Namely, if we suggest that ${\cal B}$ and ${\cal C}$ are prepared in the state $f_{\rm prep}\ket{\Phi^+}\bra{\Phi^{+}}+(1-f_{\rm pr})\rho_{\rm mix}\otimes\rho_{\rm mix}$ and the postselection effect has the form 
$f_{\rm ms}\ket{\Phi^+}\bra{\Phi^{+}}+(1-f_{\rm ms})\rho_{\rm mix}\otimes\rho_{\rm mix}$, then we obtain
\begin{equation} \label{eq:etaABCnoisy}
\begin{aligned}
	\eta_{\cal A} &= \ket{\psi}\bra{\psi} \otimes \rho_{\rm mix},\\
	\eta_{\cal B} &= \rho_{\rm mix} \otimes [f_{\rm pr}\ket{\psi}\bra{\psi} + (1-f_{\rm pr})\rho_{\rm mix}],\\
	\eta_{\cal C}  &= [f_{\rm pr}f_{\rm ms}\ket{\psi}\bra{\psi}+(1-f_{\rm pr}f_{\rm ms})\rho_{\rm mix}] \otimes \rho_{\rm mix},
\end{aligned} 
\end{equation}
where noise parameters $f_{\rm pr}$ and $f_{\rm ms}$ belong to $[0,1]$.
The form of time-bidirectional states in Eq.~\eqref{eq:etaABCnoisy} captures the decrease of fidelity and the entropy growth shown in Table~\ref{tab:rslts}.

Finally, it is worth noting that even in the presence of this kind of time travel, no logical paradoxes, such as the `grandfather paradox', appears~\cite{lloyd2011quantum, korotaev2015quantum}. 
The reason for this is the fundamental impossibility of forcing the desired postselection outcome.
In the cases of alternative Bell measurement outcomes $\ket{\Psi^+}$, $\ket{\Psi^-}$, or $\ket{\Phi^-}$, the state $\ket{\psi}$ also goes through a time-reversal trajectory, but acquires an additional unitary transformation $u$, given by $\sigma_x$, $\sigma_y$ or $\sigma_z$ correspondingly, in its `reflection' from the Bell measurement (note that this is exactly the transformation which Bob undoes after obtaining a message from Alice at the end of the quantum teleportation protocol).
Therefore, for the general postselection condition, we have
\begin{equation} \label{eq:etaABCnew}
\begin{aligned}
	\eta_{\cal A} &= \ket{\psi}\bra{\psi} \otimes \rho_{\rm mix},\\
	\eta_{\cal B} &= \rho_{\rm mix} \otimes u\ket{\psi}\bra{\psi}u^\dagger,\\
	\eta_{\cal C}  &= u\ket{\psi}\bra{\psi}u^\dagger \otimes \rho_{\rm mix}.
\end{aligned} 
\end{equation}
By removing the postselection condition completely, we arrive at 
\begin{equation} \label{eq:etaABCnops}
	\eta_{\cal A}  = \ket{\psi}\bra{\psi} \otimes \rho_{\rm mix} \quad \eta_{\cal B} = \eta_{\cal C}= \rho_{\rm mix} \otimes \rho_{\rm mix},
\end{equation}
where no any time-reversal phenomenon can be revealed.
However, we note that $\rho_{\rm mix}$ have a different physical meaning in different parts of time-bidirectional states.
In particular, $\eta_{\cal A\downarrow}=\eta_{\cal B\downarrow}=\eta_{\cal C}^\uparrow=\rho_{\rm mix}$ correspond to the uncertainty of the Bell measurement, while $\eta_{\cal B}^\downarrow=\eta_{\cal C\uparrow}=\rho_{\rm mix}$ corresponds to the uncertainty about the future if particle ${\cal C}$.

\section{Conclusion and outlook} \label{sec:concl}

In the present work, we have developed the TBSF, which unifies previously proposed time symmetrized two-state (density) vector formalism  and the standard `no postselection' formalism in general manner.
This goal is achieved by considering a generalized postselection measurement, whose particular postselection outcome is given by an arbitrary POVM effect $E_{\rm post}$.
By smoothly shifting between the limiting cases of $E_{\rm post}$ being a rank-one projector (the case of two-state vectors) and identity operator (the case of no postselection), we cover a large number of possible experimental setups, especially, where decoherence effects can not be neglected.
We have seen that the concept of a time-bidirectional state $\eta$ generalizes the concept of a quantum state $\rho$ in the standard formalism (in the no postselection case, $\eta=\rho\otimes \rho_{\rm mix}$), a concept of a two-state vector $(\ket{\psi^{\rm pre}}, \bra{\phi_{\rm post}})$ (in this case $\eta=\ket{\psi^{\rm pre}}\bra{\psi^{\rm pre}}\otimes \ket{\phi_{\rm post}}\bra{\phi_{\rm post}}$), as well as other previously considered objects such as a generalized two-state vector~\cite{aharonov1991complete}, two-state density vector~\cite{silva2014pre}, and mixed two-state vector~\cite{vaidman2017weak}.
We have derived expressions for outcome probabilities of generalized measurements, and also mean and weak values of Hermitian observables.
We have also considered practical tomography schemes for reconstructing unknown single-qubit time-bidirectional states.
Namely, we developed two schemes based MUB and SIC-POVM approaches correspondingly.
Finally, we have applied the developed formalism and the SIC-POVM-based tomography technique in order to demonstrate a time-reversal quantum state propagation in quantum teleportation on a noisy cloud-accessible superconducting processor.

The author believes that the presented formalism will be helpful for studying and developing quantum information processing protocols with postselection.
Moreover, it raises some fundamental question related to observed postselection-induced phenomena.
Can the employed formalism be for studying CTC models other from projective ones~\cite{deutsch1991quantum, shepelin2021multiworld}?
What kind of master equations can describe irreversibility on postselection-induced time arrows?
How are Markovian and non-Markovian effects different for standard evolution of a quantum state along the `macroscopic' time-arrow and postselection-induced time arrows?
Can the employed TBSF be used to study of the inherent time-asymmetry of the macroscopic world? 
What kind of effects one can expect with a `weak postselection', where $E_{\rm post}$ is close to but not equal to identity, and so on.

\section*{Acknowledgements}
The author acknowledges use of the IBM Q Experience for this work. The views expressed are those of the author and do not reflect the official policy or position of IBM or the IBM Q Experience team.
The author thanks M.A. Gavreev, I.A. Luchnikov, S.M. Korotaev, L.V. Il’ichov, A.K. Fedorov, and L. Vaidman for fruitful discussions.
The theoretical work was funded by the Russian Federation represented by the Ministry of Science and Higher Education (Grant No. 075-15-2020-788).

\appendix

\section{Cloud platform details}

Here we provide some more details about the seven-qubit cloud-accessible superconducting processor ${\sf ibm\_oslo}$, which was employed for the demonstration of time-bidirectional state tomography in Sec~\ref{sec:experiment}.
Recall, that the coupling map of the processor, corresponding to a possibility to perform CNOT gate
\begin{equation}
    {\sf CX} = \begin{bmatrix}
            1 & 0 & 0 & 0\\
            0 & 1 & 0 & 0\\
            0 & 0 & 0 & 1\\
            0 & 0 & 1 & 0
        \end{bmatrix},
\end{equation}
is show in Fig.~\ref{fig:teleporation}(c).
The set of native single qubit gates, which can be applied to any of seven qubits, consists of
\begin{equation}
    \begin{aligned}
        &{\sf ID} \equiv \mathbb{1},
        %\quad 
        &{\sf SX} \equiv R_x(\pi/2),\\
        %\quad 
        &{\sf X} \equiv R_x(\pi),
        %\quad
        &{\sf RZ}(\theta) \equiv R_z(\theta).
    \end{aligned}
\end{equation}
Calibration data at the time of the demonstration is provided in Table~\ref{tab:calibration}.

\begin{table*}[]
    \centering
    \begin{tabular}{c|c|c|c|c|c|c|c}
    Parameter & {\sf Q}0 & {\sf Q}1 & {\sf Q}2 & {\sf Q}3 & {\sf Q}4 & {\sf Q}5 & {\sf Q}6\\\hline\hline
    $T_1$ ($\mu$s) & 107.99 & 162.95 & 122.86 & 92.3 & 149.18 & 151.66 & 170.4 \\
    $T_2$ ($\mu$s) & 117.13 & 34.15 & 39.9 & 36.91 & 136.86 & 33.68 & 220.11 \\
    Frequency (GHz) & 4.925 & 5.046 & 4.962 & 5.108 & 5.011 & 5.173 & 5.319 \\
    Anharmonicity (GHz) & -0.3444 & -0.34286 & -0.34389 & -0.34119 & -0.3429 & -0.3429 & -0.33763 \\
    Readout assignment error & $8.800\times 10^{-3}$ & $1.520\times 10^{-2}$ & $9.300\times 10^{-3}$ & $1.660\times 10^{-2}$ & $2.720\times 10^{-2}$ & $1.080\times 10^{-2}$ & $3.220\times 10^{-2}$ \\
    Prob. meas 0 prep. $\ket{1}$  & 0.0118 & 0.016	 & 0.0126 & 0.0128 & 0.0264 & 0.0122 & 0.0288 \\
    Prob. meas 1 prep. $\ket{0}$  & 0.0058 & 0.0144 & 0.006	 & 0.0204 & 0.028	 & 0.0094 & 0.0356\\
    Readout length (ns)  & 910.222 & 910.222 & 910.222 & 910.222 & 910.222 & 910.222 & 910.222\\
    ID error  & $2.035\times 10^{-4}$  & $3.311\times 10^{-4}$  & $4.239\times 10^{-4}$  & $2.704\times 10^{-4}$  & $3.504\times 10^{-4}$  & $3.353\times 10^{-4}$  & $2.423\times 10^{-4}$ \\
    ${\sf SX}$ error  &	$2.035\times 10^{-4}$ &	$3.311\times 10^{-4}$ &	$4.239\times 10^{-4}$ &	$2.704\times 10^{-4}$ &	$3.504\times 10^{-4}$ &	$3.353\times 10^{-4}$ &	$2.423\times 10^{-4}$ \\
    ${\sf X}$ error  &	$2.035\times 10^{-4}$ &	$3.311\times 10^{-4}$ &	$4.239\times 10^{-4}$ &	$2.704\times 10^{-4}$ &	$3.504\times 10^{-4}$ &	$3.353\times 10^{-4}$ &	$2.423\times 10^{-4}$ \\ \hline
    \end{tabular}\\
    \vspace{5pt}{\sf CX} error\\
    \begin{tabular}{c|c|c|c|c|c|c|c}
        & {\sf Q}0 & {\sf Q}1 & {\sf Q}2 & {\sf Q}3 & {\sf Q}4 & {\sf Q}5 & {\sf Q}6 \\ \hline
         {\sf Q}0 & & $7.613\times 10^{-3}$ & & & & & \\ \hline
         {\sf Q}1 & $7.613\times 10^{-3}$ & & $9.164\times 10^{-3}$ & $6.849\times 10^{-3}$ & & & \\ \hline
         {\sf Q}2 & & $9.164\times 10^{-3}$  & & & & \\ \hline
         {\sf Q}3 & & $6.849\times 10^{-3}$ & & & &$5.382\times 10^{-3}$&  \\ \hline
         {\sf Q}4 & & & & & &  $1.090\times 10^{-2}$ &  \\ \hline
         {\sf Q}5 & & & &$5.382\times 10^{-3}$ & $1.090\times 10^{-2}$ &  &  $8.461\times 10^{-3}$  \\ \hline
         {\sf Q}6 & & & & & & $8.461\times 10^{-3}$ &
    \end{tabular}\\
    \vspace{5pt}Gate time (ns)\\
    \begin{tabular}{c|c|c|c|c|c|c|c}
        & {\sf Q}0 & {\sf Q}1 & {\sf Q}2 & {\sf Q}3 & {\sf Q}4 & {\sf Q}5 & {\sf Q}6 \\ \hline
         {\sf Q}0 & & 341.333 & & & & & \\ \hline
         {\sf Q}1 & 412.444 & & 248.889 & 263.111 & & & \\ \hline
         {\sf Q}2 & & 213.333  & & & & \\ \hline
         {\sf Q}3 & & 334.222 & & & & 238.222 &  \\ \hline
         {\sf Q}4 & & & & & &  305.778 &  \\ \hline
         {\sf Q}5 & & & &167.111 & 341.333 &  &  405.333  \\ \hline
         {\sf Q}6 & & & & & & 334.222 &
    \end{tabular}
    \caption{Calibration data for seven-qubit superconducting processor {\sf ibm\_oslo} at the time of the demonstration presented in Sec.~\ref{sec:experiment}. 
    The data is also available at~\cite{supplementary}.}
    \label{tab:calibration}
\end{table*}

\bibliography{literature.bib}

\end{document}